\documentclass[aps,preprint]{revtex4}
\usepackage{amssymb}
\usepackage{amsfonts}
\usepackage{amsmath}
\usepackage{graphicx}

\input{tcilatex}
\begin{document}

\title{Resolving game theoretical dilemmas with quantum states}
\author{Azhar Iqbal$^{\dagger }$\thanks{%
Corresponding author, Email:azhar.iqbal@adelaide.edu.au, Phone: +61427661661.%
}, James M. Chappell$^{\ddagger }$, Claudia Szabo$^{\dagger }$, Derek Abbott$%
^{\ddagger }$}
\affiliation{$^{\dagger }$School of Computer and Mathematical Sciences, University of
Adelaide, South Australia 5005, Australia\\
$^{\ddagger }$School of Electrical \& Electronic Engineering, University of
Adelaide, South Australia 5005, Australia.}

\begin{abstract}
We present a new framework for creating a quantum version of a classical
game, based on Fine's theorem. This theorem shows that for a given set of
marginals, a system of Bell's inequalities constitutes both necessary and
sufficient conditions for the existence of the corresponding joint
probability distribution. Using Fine's theorem, we reexpress both the player
payoffs and their strategies in terms of a set of marginals, thus paving the
way for the consideration of sets of marginals---corresponding to entangled
quantum states---for which no corresponding joint probability distribution
may exist. By harnessing quantum states and employing Positive
Operator-Valued Measures (POVMs), we then consider particular quantum states
that can potentially resolve dilemmas inherent in classical games.
\end{abstract}

\maketitle

\section{Introduction}

In conventional approaches \cite%
{Meyer,EWL1,EWL2,MarinattoWeber,Kolokotsov2020,GScholar} to quantizing
non-cooperative games \cite{Binmore,Osborne,Rasmusen} each player conducts
local unitary transformations (strategies) on an entangled quantum state.
Following this, a quantum measurement \cite{Peres} determines each player's
payoff. The goal is to identify sets of local unitary operations that meet
the set of Nash inequalities associated with the player payoffs. Various
methodologies and perspectives have emerged \cite%
{AlonsoSanz2019,Kolokotsov2020,FaisalKhan2018,Piotr2014,Schneider2011} to
develop quantum versions of games. Bleiler \cite{Bleiler2008} formalized a
set-theoretic approach to game quantization, introducing concepts of proper,
improper, and complete quantizations, and exploring the potential of broader
notion of randomization through the lens of quantum superposition for
quantization of games.

An entangled quantum state \cite{Peres}, when examined closely, reveals to
offer inherently quantum, non-classical probability distributions that can
breach Bell's inequalities \cite{Bell1,Bell2,Aspect,CHSH}. This
understanding inspired us to devise a framework wherein game-theoretical
challenges are articulated using the distinct attributes of quantum states,
specifically their ability to generate non-classical probability
distributions.

Fine's theorem \cite{Fine,Fine1,Fine2} asserts that a system of Bell's
inequalities, given a set of marginals, provides both necessary and
sufficient conditions to ensure the existence of a joint probability
distribution for those marginals. Building on this foundation, our paper
introduces a fresh methodology for quantum games, derived straight from the
classical game's mixed-strategy version. Harnessing Fine's theorem, we
recast each player's payffs as well as their strategies in terms of the
marginals of a joint probability distribution. Through game-theoretical
scrutiny, we can pinpoint sets of marginals that can potentially resolve the
game, subsequently enabling the identification of corresponding quantum
states. To extrapolate marginal probability distributions from these quantum
states, we delve into positive operator-valued measures (POVMs) \cite{Peres}%
---a concept previously explored in quantum game studies \cite%
{LeeJohnson2003,Kolokotsov2020}.

Our innovative approach, anchored in Fine's theorem, reframes quantum games
by focusing on Fine's theorem for a system of Bell's inequalities. By
scrutinizing the marginals of a probability distribution, our methodology
facilitates a seamless transition from a game's classical mixed-strategy
version to its quantum counterpart linked to the entangled states. This
framework offers a renewed perspective on harnessing non-classical
probability distributions, derived from entangled quantum states, to
illuminate and resolve the game-theoretical conundrums.

\section{Fine's theorem for three bivalent observables}

We consider Fine's theorem \cite{Fine,Fine1,Fine2} for three bivalent
observables that is described as follows. If $a,b,c$ are bivalent
observables (each mapping into $\left\{ +1,-1\right\} $ with given joint
distributions $P_{a,b},$ $P_{a,c},$ and $P_{b,c},$ then necessary and
sufficient for the existence of a joint distribution $P_{a,b,c},$ compatible
with the given joints for the pairs, is the satisfaction of the following
system of inequalities:

\begin{align}
P(a)+P(b)+P(c)& \leq 1+P(ab)+P(ac)+P(bc),  \notag \\
P(ab)+P(ac)& \leq P(a)+P(bc),  \notag \\
P(ab)+P(bc)& \leq P(b)+P(ac),  \notag \\
P(ac)+P(bc)& \leq P(c)+P(ab),  \label{BIs}
\end{align}%
where $P(\bullet )$ is the probability that each enclosed observable takes
the value $+1.$

In Fine's proof, when an observable $O$, for instance, takes the value $-1$
it is denoted by $\bar{O}$ iff the observable $O$ takes value $+1$, and we
also denote $\xi =P(abc)$. Then the terms in a distribution $P_{a,b,c},$ if
there were one compatible with the given joint distributions for pairs,
would have to satisfy%
\begin{eqnarray}
P(ab\bar{c}) &=&P(ab)-\xi ,  \notag \\
P(a\bar{b}c) &=&P(ac)-\xi ,  \notag \\
P(a\bar{b}\bar{c}) &=&P(a)-P(ab)-P(ac)+\xi ,  \notag \\
P(\bar{a}bc) &=&P(bc)-\xi ,  \notag \\
P(\bar{a}b\bar{c}) &=&P(b)-P(ab)-P(bc)+\xi ,  \notag \\
P(\bar{a}\bar{b}c) &=&P(c)-P(ac)-P(bc)+\xi ,  \notag \\
P(\bar{a}\bar{b}\bar{c}) &=&1-P(a)-P(b)-P(c)+P(ab)+P(ac)+P(bc)-\xi .
\label{Conditions}
\end{eqnarray}%
Using $0\leq \xi \leq \min (P(ab),P(ac),P(bc)),$ the condition that each
term in (\ref{Conditions}) be non-negative produces the system (\ref{BIs}).
Conversely, if the system (\ref{BIs}) is satisfied then choosing $\xi $ as
above insures that Eqs. (\ref{Conditions}) define the required distribution $%
P_{a,b,c}.$

\section{Applying Fine's theorem to three-player games}

Let $\Re =\left\{ \mathrm{A},\mathrm{B},\mathrm{C}\right\} $ be the set of
players. A classical three-person normal form game is given by a) Three
non-empty sets $\Sigma _{\mathrm{A}}$, $\Sigma _{\mathrm{B}}$, and $\Sigma _{%
\mathrm{C}}$ representing the strategy sets of the players \textrm{A}, 
\textrm{B}, and \textrm{C}; b) Three real valued functions $P_{\mathrm{A}}$, 
$P_{\mathrm{B}}$, and $P_{\mathrm{C}}$ defined on $\Sigma _{\mathrm{A}%
}\times \Sigma _{\mathrm{B}}\times \Sigma _{\mathrm{C}}$. The \emph{product
space} $\Sigma _{\mathrm{A}}\times \Sigma _{\mathrm{B}}\times \Sigma _{%
\mathrm{C}}$ is the set of all tuples $(\sigma _{\mathrm{A}},\sigma _{%
\mathrm{B}},\sigma _{\mathrm{C}})$ with $\sigma _{\mathrm{A}}\in \Sigma _{%
\mathrm{A}}$, $\sigma _{\mathrm{B}}\in \Sigma _{\mathrm{B}}$ and $\sigma _{%
\mathrm{C}}\in \Sigma _{\mathrm{C}}$. A strategy is understood as such a
tuple $(\sigma _{\mathrm{A}},\sigma _{\mathrm{B}},\sigma _{\mathrm{C}})$ and 
$P_{\mathrm{A}}$, $P_{\mathrm{B}}$, $P_{\mathrm{C}}$ are payoff functions of
the three players. The game is denoted as $\Gamma =\left\{ \Sigma _{\mathrm{A%
}},\Sigma _{\mathrm{B}},\Sigma _{\mathrm{C}};P_{\mathrm{A}},P_{\mathrm{B}%
},P_{\mathrm{C}}\right\} $.

To apply Fine's theorem to three-player games, we use the above notation and
consider the game for players \textrm{A}, \textrm{B}, and \textrm{C} and in
which the player payoffs are defined as

\begin{gather}
\Pi _{\mathrm{A}}(\lambda ,\mu ,\nu )=\alpha _{1}P(abc)+\beta _{1}P(ab\bar{c}%
)+\gamma _{1}P(a\bar{b}c)+\delta _{1}P(a\bar{b}\bar{c})+  \notag \\
\epsilon _{1}P(\bar{a}bc)+\varepsilon _{1}P(\bar{a}b\bar{c})+\zeta _{1}P(%
\bar{a}\bar{b}c)+\eta _{1}P(\bar{a}\bar{b}\bar{c});  \label{Payoff_A} \\
\Pi _{\mathrm{B}}(\lambda ,\mu ,\nu )=\alpha _{2}P(abc)+\beta _{2}P(ab\bar{c}%
)+\gamma _{2}P(a\bar{b}c)+\delta _{2}P(a\bar{b}\bar{c})+  \notag \\
\epsilon _{2}P(\bar{a}bc)+\varepsilon _{2}P(\bar{a}b\bar{c})+\zeta _{2}P(%
\bar{a}\bar{b}c)+\eta _{2}P(\bar{a}\bar{b}\bar{c});  \label{Payoff_B} \\
\Pi _{\mathrm{C}}(\lambda ,\mu ,\nu )=\alpha _{3}P(abc)+\beta _{3}P(ab\bar{c}%
)+\gamma _{3}P(a\bar{b}c)+\delta _{3}P(a\bar{b}\bar{c})+  \notag \\
\epsilon _{3}P(\bar{a}bc)+\varepsilon _{3}P(\bar{a}b\bar{c})+\zeta _{3}P(%
\bar{a}\bar{b}c)+\eta _{3}P(\bar{a}\bar{b}\bar{c}).  \label{Payoff_C}
\end{gather}%
where $\alpha _{1,2,3},$ $\beta _{1,2,3},$...$\eta _{1,2,3}$ are real
numbers, $\lambda ,\mu ,\nu \in \lbrack 0,1]$ and

\begin{equation}
\lambda =P(a),\text{ }\mu =P(b),\text{ }\nu =P(c),  \label{Strategies}
\end{equation}%
are defined as the (mixed) strategies of the players \textrm{A}, \textrm{B},
and \textrm{C,} respectively. Eqs. (\ref{Payoff_A},\ref{Payoff_B},\ref%
{Payoff_C}) define the player payoffs for the three players for a general
game such that it paves the way to applying Fine's theorem to the game. Note
that the observable $b$, for instance, takes the value $-1$ it is denoted by 
$\bar{b}$ iff the observable $b$ takes value $+1$. For the factorizable
case, the players' payoffs (\ref{Payoff_A},\ref{Payoff_B},\ref{Payoff_C})
are expressed as

\begin{gather}
\Pi _{\mathrm{A,B,C}}(\lambda ,\mu ,\nu )=(\alpha _{1},\alpha _{2},\alpha
_{3})\lambda \mu \nu +(\beta _{1},\beta _{2},\beta _{3})\lambda \mu (1-\nu )+
\notag \\
(\gamma _{1},\gamma _{2},\gamma _{3})\lambda (1-\mu )\nu +(\delta
_{1},\delta _{2},\delta _{3})\lambda (1-\mu )(1-\nu )+  \notag \\
(\epsilon _{1},\epsilon _{2},\epsilon _{3})(1-\lambda )\mu \nu +(\varepsilon
_{1},\varepsilon _{2},\varepsilon _{3})(1-\lambda )\mu (1-\nu )+  \notag \\
(\zeta _{1},\zeta _{2},\zeta _{3})(1-\lambda )(1-\mu )\nu +(\eta _{1},\eta
_{2},\eta _{3})(1-\lambda )(1-\mu )(1-\nu ),
\label{Factorizable_probabilities}
\end{gather}%
giving a mixed-strategy version of a three-player game.

A strategy profile is a Nash equilibrium if every strategy in that profile
is a best response to every other strategy in the profile; i.e., there is no
strategy that a player could play that would yield a higher payoff, given
all the strategies played by the other players. The Nash equilibria
described by the triple $(\lambda ^{\ast },\mu ^{\ast },\nu ^{\ast })$ for
this game, are then obtained from the inequalities

\begin{equation}
\frac{\partial \Pi _{\mathrm{A}}}{\partial \lambda }\mid _{(\lambda ^{\ast
},\mu ^{\ast },\nu ^{\ast })}(\lambda ^{\ast }-\lambda )\geq 0,\text{ }\frac{%
\partial \Pi _{\mathrm{B}}}{\partial \mu }\mid _{(\lambda ^{\ast },\mu
^{\ast },\nu ^{\ast })}(\mu ^{\ast }-\mu )\geq 0,\text{ }\frac{\partial \Pi
_{\mathrm{C}}}{\partial \nu }\mid _{(\lambda ^{\ast },\mu ^{\ast },\nu
^{\ast })}(\nu ^{\ast }-\nu )\geq 0,  \label{NIs}
\end{equation}%
for all $\lambda ,\mu ,\nu \in \lbrack 0,1].$ The inequalities (\ref{NIs})
are not sufficient as all derivatives can be zero. In that case we would
also require the second derivatives $\frac{\partial ^{2}\Pi _{\mathrm{A}}}{%
\partial \lambda ^{2}}\mid _{(\lambda ^{\ast },\mu ^{\ast },\nu ^{\ast })},$ 
$\frac{\partial ^{2}\Pi _{\mathrm{B}}}{\partial \mu ^{2}}\mid _{(\lambda
^{\ast },\mu ^{\ast },\nu ^{\ast })},$ and $\frac{\partial ^{2}\Pi _{\mathrm{%
C}}}{\partial \nu ^{2}}\mid _{(\lambda ^{\ast },\mu ^{\ast },\nu ^{\ast })}$
to be positive, in order to represent that the triple $(\lambda ^{\ast },\mu
^{\ast },\nu ^{\ast })$ is a maxima with respect to $\lambda ,\mu ,$ and $%
\nu $.

\section{The case of three-player Prisoners' Dilemma}

Consider the three-player Prisoners' Dilemma game \cite%
{Donnell1998,IqbalCheonAbbott2008} for which we write the player payoffs as

\begin{gather}
\Pi _{\mathrm{A,B,C}}(\lambda ,\mu ,\nu )=(\varkappa ,\varkappa ,\varkappa
)P(abc)+(\kappa ,\kappa ,\vartheta )P(ab\bar{c})+(\kappa ,\vartheta ,\kappa
)P(a\bar{b}c)  \notag \\
+(\tau ,\theta ,\theta )P(a\bar{b}\bar{c})+(\vartheta ,\kappa ,\kappa )P(%
\bar{a}bc)+(\theta ,\tau ,\theta )P(\bar{a}b\bar{c})  \notag \\
+(\theta ,\theta ,\tau )P(\bar{a}\bar{b}c)+(\omega ,\omega ,\omega )P(\bar{a}%
\bar{b}\bar{c}),
\end{gather}%
where

\begin{equation}
\begin{array}{l}
\text{a) }\vartheta >\varkappa ,\ \ \omega >\tau ,\ \ \theta >\kappa , \\ 
\text{b) }\vartheta >\theta >\omega ,\ \ \varkappa >\kappa >\tau , \\ 
\text{c) }\kappa >\omega ,\ \ \varkappa >\theta ,\ \ \kappa >(1/2)(\tau
+\theta ),\ \ \varkappa >(1/2)(\kappa +\vartheta ).%
\end{array}%
\end{equation}%
An example satisfying these requirements is the case when

\begin{equation}
\varkappa =7,\text{ }\vartheta =9,\text{ }\kappa =3,\text{ }\tau =0,\ \omega
=1,\text{ }\theta =5,  \label{PD_values}
\end{equation}%
for which the player payoffs (\ref{Payoff_A},\ref{Payoff_B},\ref{Payoff_C})
are expressed as

\begin{gather}
\Pi _{\mathrm{A,B,C}}(\lambda ,\mu ,\nu )=(7,7,7)P(abc)+(3,3,9)P(ab\bar{c}%
)+(3,9,3)P(a\bar{b}c)  \notag \\
+(0,5,5)P(a\bar{b}\bar{c})+(9,3,3)P(\bar{a}bc)+(5,0,5)P(\bar{a}b\bar{c}) 
\notag \\
+(5,5,0)P(\bar{a}\bar{b}c)+(1,1,1)P(\bar{a}\bar{b}\bar{c}).
\label{mixed_strategy_3_player_PD}
\end{gather}%
We have

\begin{eqnarray}
P(ab) &=&P(abc)+P(ab\bar{c})=\lambda \mu \nu +\lambda \mu (1-\nu )=\lambda
\mu ,  \notag \\
P(ac) &=&\lambda \nu ,\text{ }P(bc)=\mu \nu ,
\end{eqnarray}%
and for factorizable case, Bell's inequalities (BIs) (\ref{BIs}) then take
the form

\begin{align}
\lambda +\mu +\nu & \leq (1+\lambda \mu +\lambda \nu +\mu \nu ),  \notag \\
\lambda \mu +\lambda \nu & \leq \lambda +\mu \nu ,\text{ }\lambda \mu +\mu
\nu \leq \mu +\lambda \nu ,\text{ }\lambda \nu +\mu \nu \leq \nu +\lambda
\mu ,
\end{align}%
i.e. the sum of last three inequalities take the form

\begin{eqnarray}
2(\lambda \mu +\lambda \nu +\mu \nu ) &\leq &(\lambda +\mu +\nu )+(\lambda
\mu +\lambda \nu +\mu \nu ),  \notag \\
(\lambda \mu +\lambda \nu +\mu \nu ) &\leq &(\lambda +\mu +\nu ),  \notag \\
(\lambda \mu +\lambda \nu +\mu \nu ) &\leq &(\lambda +\mu +\nu )\leq
(1+\lambda \mu +\lambda \nu +\mu \nu ),
\end{eqnarray}%
i.e. for factorizable case, these form a consistent set as the last three
BIs are consistent with the first BI.

With (\ref{PD_values}) and for the factorizable case, the players' payoffs
are expressed as

\begin{gather}
\Pi _{\mathrm{A,B,C}}(\lambda ,\mu ,\nu )=(7,7,7)\lambda \mu \nu
+(3,3,9)\lambda \mu (1-\nu )+  \notag \\
(3,9,3)\lambda (1-\mu )\nu +(0,5,5)\lambda (1-\mu )(1-\nu )+  \notag \\
(9,3,3)(1-\lambda )\mu \nu +(5,0,5)(1-\lambda )\mu (1-\nu )+  \notag \\
(5,5,0)(1-\lambda )(1-\mu )\nu +(1,1,1)(1-\lambda )(1-\mu )(1-\nu ),
\label{PD_Factorizable}
\end{gather}%
that is

\begin{eqnarray}
\Pi _{\mathrm{A}}(\lambda ,\mu ,\nu ) &=&\lambda (\mu \nu -\mu -\nu -1)+4\mu
+4\nu +1,  \notag \\
\Pi _{\mathrm{B}}(\lambda ,\mu ,\nu ) &=&\mu (\lambda \nu -\lambda -\nu
-1)+4\nu +4\lambda +1,  \notag \\
\Pi _{\mathrm{C}}(\lambda ,\mu ,\nu ) &=&\nu (\lambda \mu -\mu -\lambda
-1)+4\mu +4\lambda +1.
\end{eqnarray}

For Nash equilibrium, as a triple $(\lambda ^{\ast },\mu ^{\ast },\nu ^{\ast
})$ for three-player PD game, we consider the inequalities (\ref{NIs}) and
obtain

\begin{eqnarray}
\frac{\partial \Pi _{\mathrm{A}}}{\partial \lambda } &\mid &_{(\lambda
^{\ast },\mu ^{\ast },\nu ^{\ast })}(\lambda ^{\ast }-\lambda )=(\mu ^{\ast
}\nu ^{\ast }-\mu ^{\ast }-\nu ^{\ast }-1)(\lambda ^{\ast }-\lambda )\geq 0,
\notag \\
\frac{\partial \Pi _{\mathrm{B}}}{\partial \mu } &\mid &_{(\lambda ^{\ast
},\mu ^{\ast },\nu ^{\ast })}(\mu ^{\ast }-\mu )=(\lambda ^{\ast }\nu ^{\ast
}-\lambda ^{\ast }-\nu ^{\ast }-1)(\mu ^{\ast }-\mu )\geq 0,  \notag \\
\frac{\partial \Pi _{\mathrm{C}}}{\partial \nu } &\mid &_{(\lambda ^{\ast
},\mu ^{\ast },\nu ^{\ast })}(\nu ^{\ast }-\nu )=(\lambda ^{\ast }\mu ^{\ast
}-\mu ^{\ast }-\lambda ^{\ast }-1)(\nu ^{\ast }-\nu )\geq 0,  \label{NIs_1}
\end{eqnarray}%
that gives the NE of $(\lambda ^{\ast },\mu ^{\ast },\nu ^{\ast })=(0,0,0)$
i.e. each of the three players play the strategy of Defection. At this NE,
the player payoffs are then obtained as $\Pi _{\mathrm{A}}(0,0,0)=\Pi _{%
\mathrm{B}}(0,0,0)=\Pi _{\mathrm{C}}(0,0,0)=1.$ Note that for strategy
triple $(1,1,1)$ each of the three players play the strategy of Cooperation.
An approach to a quantum game from the perspective of non-factorizable
probability distributions is discussed in Ref. \cite{IqbalChappellAbbott2016}
and the difference between W state and GHZ state \cite{Peres} from the
perspective of three-player PD is also discussed in Ref. \cite{HanZhang}.

\subsection{The player payoffs expressed in terms of the marginals}

We now substitute from Eqs. (\ref{Conditions},\ref{Strategies}) and the
player payoffs $\Pi _{\mathrm{A,B,C}}$ are obtained as

\begin{gather}
\Pi _{\mathrm{A,B,C}}(\lambda ,\mu ,\nu )=\alpha _{1,2,3}\xi +\beta
_{1,2,3}(P(ab)-\xi )+\gamma _{1,2,3}(P(ac)-\xi )  \notag \\
+\delta _{1,2,3}(\lambda -P(ab)-P(ac)+\xi )  \notag \\
+\epsilon _{1,2,3}(P(bc)-\xi )+\varepsilon _{1,2,3}(\mu -P(ab)-P(bc)+\xi ) 
\notag \\
+\zeta _{1,2,3}(\nu -P(ac)-P(bc)+\xi )  \notag \\
+\eta _{1,2,3}(1-\lambda -\mu -\nu +P(ab)+P(ac)+P(bc)-\xi ).  \label{payoffs}
\end{gather}%
Eqs. (\ref{payoffs}) re-express the player payoffs in terms of the
probabilities $\lambda ,\mu ,\nu $, representing the player strategies and
defined in (\ref{Strategies}). With $\xi =P(abc)$ these payoffs are
expressed as

\begin{gather}
\Pi _{\mathrm{A,B,C}}(\lambda ,\mu ,\nu )=(\alpha _{1,2,3}-\beta
_{1,2,3}-\gamma _{1,2,3}+\delta _{1,2,3}-\epsilon _{1,2,3}+\varepsilon
_{1,2,3}+\zeta _{1,2,3}-\eta _{1,2,3})P(abc)  \notag \\
+(\beta _{1,2,3}-\delta _{1,2,3}-\varepsilon _{1,2,3}+\eta
_{1,2,3})P(ab)+(\epsilon _{1,2,3}-\varepsilon _{1,2,3}-\zeta _{1,2,3}+\eta
_{1,2,3})P(bc)  \notag \\
+(\gamma _{1,2,3}-\delta _{1,2,3}-\zeta _{1,2,3}+\eta _{1,2,3})P(ac)+(\delta
_{1,2,3}-\eta _{1,2,3})\lambda  \notag \\
+(\varepsilon _{1,2,3}-\eta _{1,2,3})\mu +(\zeta _{1,2,3}-\eta _{1,2,3})\nu
+\eta _{1,2,3}.  \label{payoffs_1}
\end{gather}

We consider Eqs. (\ref{payoffs_1}) as the defining payoff relations for the
corresponding quantum game, as these payoffs also allow us to consider the
case when Bell's inequalities (\ref{BIs}) are violated. This will be the
case when the distribution $P_{a,b,c}$ as defined in (\ref{Conditions}) does
not hold.

We emphasize here that when Bell's inequalities hold, the steps in obtaining
Eqs. (\ref{payoffs_1}) from (\ref{Payoff_A},\ref{Payoff_B},\ref{Payoff_C})
can be reversed. This is because Bell's inequalities (\ref{BIs}) are
necessary and sufficient for the existence of a joint distribution $\xi $.
That is, from the given payoff relations (\ref{payoffs_1}), one can obtain
the payoff relations (\ref{Payoff_A},\ref{Payoff_B},\ref{Payoff_C}) from
which the classical mixed-strategy game can then be obtained for the case of
factorizable probabilities as considered in Eq. (\ref%
{Factorizable_probabilities}).

\subsection{Expressing the players' payoffs in three-player Prisoners'
Dilemma in terms of the marginals}

After substituting from Eqs. (\ref{PD_values},\ref{Conditions}, \ref%
{Strategies}), the player payoffs $\Pi _{\mathrm{A,B,C}}$ above are obtained
as

\begin{gather}
\Pi _{\mathrm{A}}(\lambda ,\mu ,\nu )=\xi -P(ab)-P(ac)-\lambda +4\mu +4\nu
+1,  \notag \\
\Pi _{\mathrm{B}}(\lambda ,\mu ,\nu )=\xi -P(ab)-P(bc)+4\lambda -\mu +4\nu
+1,  \notag \\
\Pi _{\mathrm{C}}(\lambda ,\mu ,\nu )=\xi -P(bc)-P(ac)+4\lambda +4\mu -\nu
+1.  \label{PD_GHZ}
\end{gather}

Now, if Bell's inequalities hold then the joint distribution (\ref%
{Conditions}) exists and the player payoffs (\ref{PD_GHZ}) take the form of
those in the classical mixed-strategy game.

\section{Using quantum states}

We now use a quantum state $\left\vert \Psi \right\rangle $ and assume that
the bivalent observables $a,b,$ and $c$ are the eigenvalues of Pauli
spin-flip operator $\sigma _{z}$. The diagonal representation of $\sigma
_{z} $ is $\sigma _{z}=\left\vert 0\right\rangle \left\langle 0\right\vert
-\left\vert 1\right\rangle \left\langle 1\right\vert $ and with $\left\vert
0\right\rangle =\left( 
\begin{array}{c}
1 \\ 
0%
\end{array}%
\right) $ and $\left\vert 1\right\rangle =\left( 
\begin{array}{c}
0 \\ 
1%
\end{array}%
\right) $ the eigenstates of $\sigma _{z}$ are $\left\vert \psi
\right\rangle _{+1}=\left\vert 0\right\rangle $ and $\left\vert \psi
\right\rangle _{-1}=\left\vert 1\right\rangle $ with the eigenvalues $+1$
and $-1$, respectively. Let $\rho _{\Psi }$ be the density matrix of the
quantum state $\left\vert \Psi \right\rangle $.

\subsection{Evaluating the marginals in Bell's inequalities using POVMs}

A positive operator-valued measure (POVM) \cite{Peres} is a measure whose
values are positive semi-definite operators on a Hilbert space and the
quantum measurements described by POVMs are a generalization of quantum
measurement. To measure Alice's system in the standard basis, we consider
the POVM given by

\begin{equation}
M_{\mathrm{A}}^{0}=\left\vert 0\right\rangle \left\langle 0\right\vert _{%
\mathrm{A}}\otimes \mathcal{I}_{\mathrm{B}}\otimes \mathcal{I}_{\mathrm{C}},%
\text{ }M_{\mathrm{A}}^{1}=\left\vert 1\right\rangle \left\langle
1\right\vert _{\mathrm{A}}\otimes \mathcal{I}_{\mathrm{B}}\otimes \mathcal{I}%
_{\mathrm{C}},
\end{equation}%
for which we have

\begin{equation}
M_{\mathrm{A}}^{0}+M_{\mathrm{A}}^{1}=(\left\vert 0\right\rangle
\left\langle 0\right\vert _{\mathrm{A}}+\left\vert 1\right\rangle
\left\langle 1\right\vert _{\mathrm{A}})\otimes \mathcal{I}_{\mathrm{B}%
}\otimes \mathcal{I}_{\mathrm{C}}=\mathcal{I}_{\mathrm{A}}\otimes \mathcal{I}%
_{\mathrm{B}}\otimes \mathcal{I}_{\mathrm{C}}=\mathcal{I},
\end{equation}%
and similarly%
\begin{equation}
M_{\mathrm{B}}^{0}+M_{\mathrm{B}}^{1}=M_{\mathrm{C}}^{0}+M_{\mathrm{C}}^{1}=%
\mathcal{I}.
\end{equation}%
The probability of Alice, Bob, and Chris obtaining the outcome $+1$,
respectively, is then obtained as

\begin{equation}
\lambda =P(a)=\mathrm{Tr}(M_{\mathrm{A}}^{0}\rho ),\text{ }\mu =P(b)=\mathrm{%
Tr}(M_{\mathrm{B}}^{0}\rho ),\text{ }\nu =P(c)=\mathrm{Tr}(M_{\mathrm{C}%
}^{0}\rho ).  \label{strategies_as_traces}
\end{equation}%
For the probability $P(ab)$ we consider the POVM given by

\begin{eqnarray}
M_{\mathrm{AB}}^{00} &=&\left\vert 0\right\rangle \left\langle 0\right\vert
_{\mathrm{A}}\otimes \left\vert 0\right\rangle \left\langle 0\right\vert _{%
\mathrm{B}}\otimes \mathcal{I}_{\mathrm{C}},\text{ }M_{\mathrm{AB}%
}^{11}=\left\vert 1\right\rangle \left\langle 1\right\vert _{\mathrm{A}%
}\otimes \left\vert 1\right\rangle \left\langle 1\right\vert _{\mathrm{B}%
}\otimes \mathcal{I}_{\mathrm{C}},  \notag \\
M_{\mathrm{AB}}^{10} &=&\left\vert 1\right\rangle \left\langle 1\right\vert
_{\mathrm{A}}\otimes \left\vert 0\right\rangle \left\langle 0\right\vert _{%
\mathrm{B}}\otimes \mathcal{I}_{\mathrm{C}},\text{ }M_{\mathrm{AB}%
}^{01}=\left\vert 0\right\rangle \left\langle 0\right\vert _{\mathrm{A}%
}\otimes \left\vert 1\right\rangle \left\langle 1\right\vert _{\mathrm{B}%
}\otimes \mathcal{I}_{\mathrm{C}}.
\end{eqnarray}%
Note that

\begin{gather}
M_{\mathrm{AB}}^{00}+M_{\mathrm{AB}}^{01}=(\left\vert 0\right\rangle
\left\langle 0\right\vert _{\mathrm{A}}\otimes \left\vert 0\right\rangle
\left\langle 0\right\vert _{\mathrm{B}}+\left\vert 0\right\rangle
\left\langle 0\right\vert _{\mathrm{A}}\otimes \left\vert 1\right\rangle
\left\langle 1\right\vert _{\mathrm{B}})\otimes \mathcal{I}_{\mathrm{C}} 
\notag \\
=\left\vert 0\right\rangle \left\langle 0\right\vert _{\mathrm{A}}\otimes
(\left\vert 0\right\rangle \left\langle 0\right\vert _{\mathrm{B}%
}+\left\vert 1\right\rangle \left\langle 1\right\vert _{\mathrm{B}})\otimes 
\mathcal{I}_{\mathrm{C}}  \notag \\
=\left\vert 0\right\rangle \left\langle 0\right\vert _{\mathrm{A}}\otimes 
\mathcal{I}_{\mathrm{B}}\otimes \mathcal{I}_{\mathrm{C}},\text{ and}  \notag
\\
M_{\mathrm{AB}}^{11}+M_{\mathrm{AB}}^{10}=(\left\vert 1\right\rangle
\left\langle 1\right\vert _{\mathrm{A}}\otimes \left\vert 1\right\rangle
\left\langle 1\right\vert _{\mathrm{B}}+\left\vert 1\right\rangle
\left\langle 1\right\vert _{\mathrm{A}}\otimes \left\vert 0\right\rangle
\left\langle 0\right\vert _{\mathrm{B}})\otimes \mathcal{I}_{\mathrm{C}} 
\notag \\
=\left\vert 1\right\rangle \left\langle 1\right\vert _{\mathrm{A}}\otimes
(\left\vert 1\right\rangle \left\langle 1\right\vert _{\mathrm{B}%
}+\left\vert 0\right\rangle \left\langle 0\right\vert _{\mathrm{B}})\otimes 
\mathcal{I}_{\mathrm{C}}  \notag \\
=\left\vert 1\right\rangle \left\langle 1\right\vert _{\mathrm{A}}\otimes 
\mathcal{I}_{\mathrm{B}}\otimes \mathcal{I}_{\mathrm{C}},\text{ and,
therefore}  \notag \\
M_{\mathrm{AB}}^{00}+M_{\mathrm{AB}}^{01}+M_{\mathrm{AB}}^{11}+M_{\mathrm{AB}%
}^{10}=(\left\vert 0\right\rangle \left\langle 0\right\vert _{\mathrm{A}%
}+\left\vert 1\right\rangle \left\langle 1\right\vert _{\mathrm{A}})\otimes 
\mathcal{I}_{\mathrm{B}}\otimes \mathcal{I}_{\mathrm{C}}  \notag \\
=\mathcal{I}_{\mathrm{A}}\otimes \mathcal{I}_{\mathrm{B}}\otimes \mathcal{I}%
_{\mathrm{C}}=\mathcal{I}
\end{gather}%
We can define another POVM given by

\begin{eqnarray}
M_{\mathrm{AB}}^{+1} &=&M_{\mathrm{AB}}^{00}+M_{\mathrm{AB}}^{11}  \notag \\
&=&(\left\vert 0\right\rangle \left\langle 0\right\vert _{\mathrm{A}}\otimes
\left\vert 0\right\rangle \left\langle 0\right\vert _{\mathrm{B}}+\left\vert
1\right\rangle \left\langle 1\right\vert _{\mathrm{A}}\otimes \left\vert
1\right\rangle \left\langle 1\right\vert _{\mathrm{B}})\otimes \mathcal{I}_{%
\mathrm{C}}, \\
M_{\mathrm{AB}}^{-1} &=&M_{\mathrm{AB}}^{10}+M_{\mathrm{AB}}^{01}  \notag \\
&=&(\left\vert 1\right\rangle \left\langle 1\right\vert _{\mathrm{A}}\otimes
\left\vert 0\right\rangle \left\langle 0\right\vert _{\mathrm{B}}+\left\vert
0\right\rangle \left\langle 0\right\vert _{\mathrm{A}}\otimes \left\vert
1\right\rangle \left\langle 1\right\vert _{\mathrm{B}})\otimes \mathcal{I}_{%
\mathrm{C}},
\end{eqnarray}%
and we obtain $P(ab)=\mathrm{Tr}(M_{\mathrm{AB}}^{+1}\rho ).$

Similarly, for $P(bc)$ we consider the POVM given by

\begin{eqnarray}
M_{\mathrm{BC}}^{00} &=&\mathcal{I}_{\mathrm{A}}\otimes (\left\vert
0\right\rangle \left\langle 0\right\vert _{\mathrm{B}}\otimes \left\vert
0\right\rangle \left\langle 0\right\vert _{\mathrm{C}}),\text{ }M_{\mathrm{BC%
}}^{11}=\mathcal{I}_{\mathrm{A}}\otimes (\left\vert 1\right\rangle
\left\langle 1\right\vert _{\mathrm{B}}\otimes \left\vert 1\right\rangle
\left\langle 1\right\vert _{\mathrm{C}}),  \notag \\
M_{\mathrm{BC}}^{10} &=&\mathcal{I}_{\mathrm{A}}\otimes (\left\vert
1\right\rangle \left\langle 1\right\vert _{\mathrm{B}}\otimes \left\vert
0\right\rangle \left\langle 0\right\vert _{\mathrm{C}}),\text{ }M_{\mathrm{BC%
}}^{01}=\mathcal{I}_{\mathrm{A}}\otimes (\left\vert 0\right\rangle
\left\langle 0\right\vert _{\mathrm{B}}\otimes \left\vert 1\right\rangle
\left\langle 1\right\vert _{\mathrm{C}}).
\end{eqnarray}

\begin{eqnarray}
M_{\mathrm{BC}}^{+1} &=&M_{\mathrm{BC}}^{00}+M_{\mathrm{BC}}^{11}  \notag \\
&=&\mathcal{I}_{\mathrm{A}}\otimes (\left\vert 0\right\rangle \left\langle
0\right\vert _{\mathrm{B}}\otimes \left\vert 0\right\rangle \left\langle
0\right\vert _{\mathrm{C}}+\left\vert 1\right\rangle \left\langle
1\right\vert _{\mathrm{B}}\otimes \left\vert 1\right\rangle \left\langle
1\right\vert _{\mathrm{C}}), \\
M_{\mathrm{BC}}^{-1} &=&M_{\mathrm{BC}}^{10}+M_{\mathrm{BC}}^{01}  \notag \\
&=&\mathcal{I}_{\mathrm{A}}\otimes (\left\vert 1\right\rangle \left\langle
1\right\vert _{\mathrm{B}}\otimes \left\vert 0\right\rangle \left\langle
0\right\vert _{\mathrm{C}}+\left\vert 0\right\rangle \left\langle
0\right\vert _{\mathrm{B}}\otimes \left\vert 1\right\rangle \left\langle
1\right\vert _{\mathrm{C}}),
\end{eqnarray}%
and we obtain $P(bc)=\mathrm{Tr}(M_{\mathrm{BC}}^{+1}\rho ).$

Similarly, for $P(ac)$ we consider the POVM given by

\begin{eqnarray}
M_{\mathrm{AC}}^{00} &=&\left\vert 0\right\rangle \left\langle 0\right\vert
_{\mathrm{A}}\otimes \mathcal{I}_{\mathrm{B}}\otimes \left\vert
0\right\rangle \left\langle 0\right\vert _{\mathrm{C}},\text{ }M_{\mathrm{AC}%
}^{11}=\left\vert 1\right\rangle \left\langle 1\right\vert _{\mathrm{A}%
}\otimes \mathcal{I}_{\mathrm{B}}\otimes \left\vert 1\right\rangle
\left\langle 1\right\vert _{\mathrm{C}},  \notag \\
M_{\mathrm{AC}}^{10} &=&\left\vert 1\right\rangle \left\langle 1\right\vert
_{\mathrm{A}}\otimes \mathcal{I}_{\mathrm{B}}\otimes \left\vert
0\right\rangle \left\langle 0\right\vert _{\mathrm{C}},\text{ }M_{\mathrm{AC}%
}^{01}=\left\vert 0\right\rangle \left\langle 0\right\vert _{\mathrm{A}%
}\otimes \mathcal{I}_{\mathrm{B}}\otimes \left\vert 1\right\rangle
\left\langle 1\right\vert _{\mathrm{C}},
\end{eqnarray}%
and we note that

\begin{eqnarray}
M_{\mathrm{AC}}^{+1} &=&M_{\mathrm{AC}}^{00}+M_{\mathrm{AC}}^{11}  \notag \\
&=&\left\vert 0\right\rangle \left\langle 0\right\vert _{\mathrm{A}}\otimes 
\mathcal{I}_{\mathrm{B}}\otimes \left\vert 0\right\rangle \left\langle
0\right\vert _{\mathrm{C}}+\left\vert 1\right\rangle \left\langle
1\right\vert _{\mathrm{A}}\otimes \mathcal{I}_{\mathrm{B}}\otimes \left\vert
1\right\rangle \left\langle 1\right\vert _{\mathrm{C}}, \\
M_{\mathrm{AC}}^{-1} &=&M_{\mathrm{AC}}^{10}+M_{\mathrm{AC}}^{01}  \notag \\
&=&\left\vert 1\right\rangle \left\langle 1\right\vert _{\mathrm{A}}\otimes 
\mathcal{I}_{\mathrm{B}}\otimes \left\vert 0\right\rangle \left\langle
0\right\vert _{\mathrm{C}}+\left\vert 0\right\rangle \left\langle
0\right\vert _{\mathrm{A}}\otimes \mathcal{I}_{\mathrm{B}}\otimes \left\vert
1\right\rangle \left\langle 1\right\vert _{\mathrm{C}},
\end{eqnarray}%
and we obtain $P(ac)=\mathrm{Tr}(M_{\mathrm{AC}}^{+1}\rho ).$ Note that

\begin{eqnarray}
M_{\mathrm{BC}}^{00} &=&(\left\vert 0\right\rangle \left\langle 0\right\vert
_{\mathrm{A}}+\left\vert 1\right\rangle \left\langle 1\right\vert _{\mathrm{A%
}})\otimes (\left\vert 0\right\rangle \left\langle 0\right\vert _{\mathrm{B}%
}\otimes \left\vert 0\right\rangle \left\langle 0\right\vert _{\mathrm{C}}) 
\notag \\
&=&\left\vert 0\right\rangle \left\langle 0\right\vert _{\mathrm{A}}\otimes
\left\vert 0\right\rangle \left\langle 0\right\vert _{\mathrm{B}}\otimes
\left\vert 0\right\rangle \left\langle 0\right\vert _{\mathrm{C}}+\left\vert
1\right\rangle \left\langle 1\right\vert _{\mathrm{A}}\otimes \left\vert
0\right\rangle \left\langle 0\right\vert _{\mathrm{B}}\otimes \left\vert
0\right\rangle \left\langle 0\right\vert _{\mathrm{C}}, \\
M_{\mathrm{BC}}^{11} &=&(\left\vert 0\right\rangle \left\langle 0\right\vert
_{\mathrm{A}}+\left\vert 1\right\rangle \left\langle 1\right\vert _{\mathrm{A%
}})\otimes (\left\vert 1\right\rangle \left\langle 1\right\vert _{\mathrm{B}%
}\otimes \left\vert 1\right\rangle \left\langle 1\right\vert _{\mathrm{C}}) 
\notag \\
&=&\left\vert 0\right\rangle \left\langle 0\right\vert _{\mathrm{A}}\otimes
\left\vert 1\right\rangle \left\langle 1\right\vert _{\mathrm{B}}\otimes
\left\vert 1\right\rangle \left\langle 1\right\vert _{\mathrm{C}}+\left\vert
1\right\rangle \left\langle 1\right\vert _{\mathrm{A}}\otimes \left\vert
1\right\rangle \left\langle 1\right\vert _{\mathrm{B}}\otimes \left\vert
1\right\rangle \left\langle 1\right\vert _{\mathrm{C}}, \\
M_{\mathrm{BC}}^{10} &=&(\left\vert 0\right\rangle \left\langle 0\right\vert
_{\mathrm{A}}+\left\vert 1\right\rangle \left\langle 1\right\vert _{\mathrm{A%
}})\otimes (\left\vert 1\right\rangle \left\langle 1\right\vert _{\mathrm{B}%
}\otimes \left\vert 0\right\rangle \left\langle 0\right\vert _{\mathrm{C}}) 
\notag \\
&=&\left\vert 0\right\rangle \left\langle 0\right\vert _{\mathrm{A}}\otimes
\left\vert 1\right\rangle \left\langle 1\right\vert _{\mathrm{B}}\otimes
\left\vert 0\right\rangle \left\langle 0\right\vert _{\mathrm{C}}+\left\vert
1\right\rangle \left\langle 1\right\vert _{\mathrm{A}}\otimes \left\vert
1\right\rangle \left\langle 1\right\vert _{\mathrm{B}}\otimes \left\vert
0\right\rangle \left\langle 0\right\vert _{\mathrm{C}}, \\
M_{\mathrm{BC}}^{01} &=&(\left\vert 0\right\rangle \left\langle 0\right\vert
_{\mathrm{A}}+\left\vert 1\right\rangle \left\langle 1\right\vert _{\mathrm{A%
}})\otimes (\left\vert 0\right\rangle \left\langle 0\right\vert _{\mathrm{B}%
}\otimes \left\vert 1\right\rangle \left\langle 1\right\vert _{\mathrm{C}}) 
\notag \\
&=&\left\vert 0\right\rangle \left\langle 0\right\vert _{\mathrm{A}}\otimes
\left\vert 0\right\rangle \left\langle 0\right\vert _{\mathrm{B}}\otimes
\left\vert 1\right\rangle \left\langle 1\right\vert _{\mathrm{C}}+\left\vert
1\right\rangle \left\langle 1\right\vert _{\mathrm{A}}\otimes \left\vert
0\right\rangle \left\langle 0\right\vert _{\mathrm{B}}\otimes \left\vert
1\right\rangle \left\langle 1\right\vert _{\mathrm{C}}.
\end{eqnarray}%
Now consider

\begin{eqnarray}
M_{\mathrm{ABC}}^{+1} &=&\left\vert 0\right\rangle \left\langle 0\right\vert
_{\mathrm{A}}\otimes \left\vert 0\right\rangle \left\langle 0\right\vert _{%
\mathrm{B}}\otimes \left\vert 0\right\rangle \left\langle 0\right\vert _{%
\mathrm{C}}+\left\vert 1\right\rangle \left\langle 1\right\vert _{\mathrm{A}%
}\otimes \left\vert 1\right\rangle \left\langle 1\right\vert _{\mathrm{B}%
}\otimes \left\vert 0\right\rangle \left\langle 0\right\vert _{\mathrm{C}}+ 
\notag \\
&&\left\vert 0\right\rangle \left\langle 0\right\vert _{\mathrm{A}}\otimes
\left\vert 1\right\rangle \left\langle 1\right\vert _{\mathrm{B}}\otimes
\left\vert 1\right\rangle \left\langle 1\right\vert _{\mathrm{C}}+\left\vert
1\right\rangle \left\langle 1\right\vert _{\mathrm{A}}\otimes \left\vert
0\right\rangle \left\langle 0\right\vert _{\mathrm{B}}\otimes \left\vert
1\right\rangle \left\langle 1\right\vert _{\mathrm{C}}, \\
M_{\mathrm{ABC}}^{-1} &=&\left\vert 1\right\rangle \left\langle 1\right\vert
_{\mathrm{A}}\otimes \left\vert 1\right\rangle \left\langle 1\right\vert _{%
\mathrm{B}}\otimes \left\vert 1\right\rangle \left\langle 1\right\vert _{%
\mathrm{C}}+\left\vert 1\right\rangle \left\langle 1\right\vert _{\mathrm{A}%
}\otimes \left\vert 0\right\rangle \left\langle 0\right\vert _{\mathrm{B}%
}\otimes \left\vert 0\right\rangle \left\langle 0\right\vert _{\mathrm{C}}+ 
\notag \\
&&\left\vert 0\right\rangle \left\langle 0\right\vert _{\mathrm{A}}\otimes
\left\vert 1\right\rangle \left\langle 1\right\vert _{\mathrm{B}}\otimes
\left\vert 0\right\rangle \left\langle 0\right\vert _{\mathrm{C}}+\left\vert
0\right\rangle \left\langle 0\right\vert _{\mathrm{A}}\otimes \left\vert
0\right\rangle \left\langle 0\right\vert _{\mathrm{B}}\otimes \left\vert
1\right\rangle \left\langle 1\right\vert _{\mathrm{C}},
\end{eqnarray}%
for which

\begin{equation}
M_{\mathrm{ABC}}^{+1}+M_{\mathrm{ABC}}^{-1}=(\left\vert 0\right\rangle
\left\langle 0\right\vert _{\mathrm{A}}\otimes \left\vert 1\right\rangle
\left\langle 1\right\vert _{\mathrm{A}})\otimes \mathcal{I}_{\mathrm{B}%
}\otimes \mathcal{I}_{\mathrm{C}}=\mathcal{I}_{\mathrm{A}}\otimes \mathcal{I}%
_{\mathrm{B}}\otimes \mathcal{I}_{\mathrm{C}}=\mathcal{I}
\end{equation}%
That is, $M_{\mathrm{ABC}}^{+1},$ $M_{\mathrm{ABC}}^{-1}$ form a POVM.

\subsection{Obtaining the marginals using POVMs}

We consider GHZ type pure state i.e. $\left\vert \Psi \right\rangle
=a\left\vert 000\right\rangle +b\left\vert 111\right\rangle $ where $a,b\in $
$\mathbb{C}$ and $\left\vert a\right\vert ^{2}+\left\vert b\right\vert
^{2}=1 $. This gives $\left\langle \Psi \right\vert =a^{\ast }\left\langle
000\right\vert +b^{\ast }\left\langle 111\right\vert $ and we have

\begin{equation}
\rho _{\Psi }=\left\vert \Psi \right\rangle \left\langle \Psi \right\vert
=\left\vert a\right\vert ^{2}\left\vert 000\right\rangle \left\langle
000\right\vert +ab^{\ast }\left\vert 000\right\rangle \left\langle
111\right\vert +a^{\ast }b\left\vert 111\right\rangle \left\langle
000\right\vert +\left\vert b\right\vert ^{2}\left\vert 111\right\rangle
\left\langle 111\right\vert ,
\end{equation}%
and from (\ref{strategies_as_traces}) we therefore have

\begin{equation}
\lambda =P(a)=\mathrm{Tr}(M_{\mathrm{A}}^{0}\rho _{\Psi }),\mu =P(b)=\mathrm{%
Tr}(M_{\mathrm{B}}^{0}\rho _{\Psi }),\nu =P(c)=\mathrm{Tr}(M_{\mathrm{C}%
}^{0}\rho _{\Psi }).
\end{equation}

\begin{eqnarray}
\lambda &=&\mathrm{Tr}(\mathrm{M_{\mathrm{A}}^{0}\rho _{\Psi }})=\mathrm{Tr}%
\left[ (\left\vert 0\right\rangle \left\langle 0\right\vert _{\mathrm{A}%
}\otimes \mathcal{I}_{\mathrm{B}}\otimes \mathcal{I}_{\mathrm{C}})\mathrm{%
\rho _{\Psi }}\right]  \notag \\
&=&\mathrm{Tr}[\left\vert a\right\vert ^{2}(\left\vert 000\right\rangle
\left\langle 000\right\vert +ab^{\ast }\left\vert 000\right\rangle
\left\langle 111\right\vert )=\left\vert a\right\vert ^{2}.
\end{eqnarray}%
and

\begin{eqnarray}
\mu &=&\mathrm{Tr}(\mathrm{M_{\mathrm{B}}^{0}\rho _{\Psi }})=\mathrm{Tr}%
\left[ (\mathcal{I}_{\mathrm{A}}\otimes \left\vert 0\right\rangle
\left\langle 0\right\vert _{\mathrm{B}}\otimes \mathcal{I}_{\mathrm{C}})%
\mathrm{\rho _{\Psi }}\right]  \notag \\
&=&\mathrm{Tr}[\left\vert a\right\vert ^{2}(\left\vert 000\right\rangle
\left\langle 000\right\vert +ab^{\ast }\left\vert 000\right\rangle
\left\langle 111\right\vert )=\left\vert a\right\vert ^{2}.
\end{eqnarray}%
and

\begin{eqnarray}
\nu &=&\mathrm{Tr}(\mathrm{M_{\mathrm{C}}^{0}\rho _{\Psi }})=\mathrm{Tr}%
\left[ (\mathcal{I}_{\mathrm{A}}\otimes \mathcal{I}_{\mathrm{B}}\otimes
\left\vert 0\right\rangle \left\langle 0\right\vert _{\mathrm{C}})\mathrm{%
\rho _{\Psi }}\right]  \notag \\
&=&\mathrm{Tr}[\left\vert a\right\vert ^{2}(\left\vert 000\right\rangle
\left\langle 000\right\vert +ab^{\ast }\left\vert 000\right\rangle
\left\langle 111\right\vert )=\left\vert a\right\vert ^{2}.
\end{eqnarray}

Also, as $P(ab)=\mathrm{Tr}(M_{\mathrm{AB}}^{+1}\mathrm{\rho _{\Psi }}),$ we
obtain

\begin{gather}
P(ab)=\mathrm{Tr}[(\left\vert 0\right\rangle \left\langle 0\right\vert _{%
\mathrm{A}}\otimes \left\vert 0\right\rangle \left\langle 0\right\vert _{%
\mathrm{B}}+\left\vert 1\right\rangle \left\langle 1\right\vert _{\mathrm{A}%
}\otimes \left\vert 1\right\rangle \left\langle 1\right\vert _{\mathrm{B}%
})\otimes \mathcal{I}_{\mathrm{C}}]\mathrm{\rho _{\Psi }}  \notag \\
=\mathrm{Tr}(\left\vert a\right\vert ^{2}\left\vert 000\right\rangle
\left\langle 000\right\vert +ab^{\ast }\left\vert 000\right\rangle
\left\langle 111\right\vert +a^{\ast }b\left\vert 111\right\rangle
\left\langle 000\right\vert +\left\vert b\right\vert ^{2}\left\vert
111\right\rangle \left\langle 111\right\vert )  \notag \\
=\left\vert a\right\vert ^{2}+\left\vert b\right\vert ^{2}=1
\end{gather}%
and we obtain $P(ab)=1$.

Similarly, as $P(bc)=\mathrm{Tr}(M_{\mathrm{BC}}^{+1}\mathrm{\rho _{\Psi }}%
), $ we obtain

\begin{eqnarray}
\mathrm{Tr}(M_{\mathrm{BC}}^{+1}\mathrm{\rho _{\Psi }}) &=&\mathrm{Tr}[%
\mathcal{I}_{\mathrm{A}}\otimes (\left\vert 0\right\rangle \left\langle
0\right\vert _{\mathrm{B}}\otimes \left\vert 0\right\rangle \left\langle
0\right\vert _{\mathrm{C}}+\left\vert 1\right\rangle \left\langle
1\right\vert _{\mathrm{B}}\otimes \left\vert 1\right\rangle \left\langle
1\right\vert _{\mathrm{C}})]\mathrm{\rho _{\Psi }}  \notag \\
&=&\mathrm{Tr}(\left\vert a\right\vert ^{2}\left\vert 000\right\rangle
\left\langle 000\right\vert +ab^{\ast }\left\vert 000\right\rangle
\left\langle 111\right\vert +a^{\ast }b\left\vert 111\right\rangle
\left\langle 000\right\vert +\left\vert b\right\vert ^{2}\left\vert
111\right\rangle \left\langle 111\right\vert )  \notag \\
&=&\left\vert a\right\vert ^{2}+\left\vert b\right\vert ^{2}=1,
\end{eqnarray}%
and, as $P(ac)=\mathrm{Tr}(M_{\mathrm{AC}}^{+1}\mathrm{\rho _{\Psi }}),$ we
obtain

\begin{eqnarray}
\mathrm{Tr}(M_{\mathrm{AC}}^{+1}\mathrm{\rho _{\Psi }}) &=&\mathrm{Tr}%
[\left\vert 0\right\rangle \left\langle 0\right\vert _{\mathrm{A}}\otimes 
\mathcal{I}_{\mathrm{B}}\otimes \left\vert 0\right\rangle \left\langle
0\right\vert _{\mathrm{C}}+\left\vert 1\right\rangle \left\langle
1\right\vert _{\mathrm{A}}\otimes \mathcal{I}_{\mathrm{B}}\otimes \left\vert
1\right\rangle \left\langle 1\right\vert _{\mathrm{C}}]\mathrm{\rho _{\Psi }}
\notag \\
&=&\mathrm{Tr}(\left\vert a\right\vert ^{2}\left\vert 000\right\rangle
\left\langle 000\right\vert +ab^{\ast }\left\vert 000\right\rangle
\left\langle 111\right\vert +a^{\ast }b\left\vert 111\right\rangle
\left\langle 000\right\vert +\left\vert b\right\vert ^{2}\left\vert
111\right\rangle \left\langle 111\right\vert )  \notag \\
&=&\left\vert a\right\vert ^{2}+\left\vert b\right\vert ^{2}=1
\end{eqnarray}%
and we obtain $P(ac)=1.$ Similarly, as $P(abc)=\mathrm{Tr}(M_{\mathrm{ABC}%
}^{+1}\mathrm{\rho _{\Psi }}),$ we obtain

\begin{eqnarray}
\mathrm{Tr}(M_{\mathrm{ABC}}^{+1}\mathrm{\rho _{\Psi }}) &=&\mathrm{Tr}%
[\left\vert 0\right\rangle \left\langle 0\right\vert _{\mathrm{A}}\otimes
\left\vert 0\right\rangle \left\langle 0\right\vert _{\mathrm{B}}\otimes
\left\vert 0\right\rangle \left\langle 0\right\vert _{\mathrm{C}}+\left\vert
1\right\rangle \left\langle 1\right\vert _{\mathrm{A}}\otimes \left\vert
1\right\rangle \left\langle 1\right\vert _{\mathrm{B}}\otimes \left\vert
0\right\rangle \left\langle 0\right\vert _{\mathrm{C}}+  \notag \\
&&\left\vert 0\right\rangle \left\langle 0\right\vert _{\mathrm{A}}\otimes
\left\vert 1\right\rangle \left\langle 1\right\vert _{\mathrm{B}}\otimes
\left\vert 1\right\rangle \left\langle 1\right\vert _{\mathrm{C}}+\left\vert
1\right\rangle \left\langle 1\right\vert _{\mathrm{A}}\otimes \left\vert
0\right\rangle \left\langle 0\right\vert _{\mathrm{B}}\otimes \left\vert
1\right\rangle \left\langle 1\right\vert _{\mathrm{C}}]\mathrm{\rho _{\Psi }}
\notag \\
&=&\mathrm{Tr}(\left\vert a\right\vert ^{2}\left\vert 000\right\rangle
\left\langle 000\right\vert +ab^{\ast }\left\vert 000\right\rangle
\left\langle 111\right\vert )=\left\vert a\right\vert ^{2},
\end{eqnarray}%
and we obtain $P(abc)=\left\vert a\right\vert ^{2}$ and Bell's inequalities (%
\ref{BIs}) take the form

\begin{equation}
3\left\vert a\right\vert ^{2}\leq 4,\text{ }2\leq \left\vert a\right\vert
^{2}+1,\text{ }2\leq \left\vert a\right\vert ^{2}+1,\text{ }2\leq \left\vert
a\right\vert ^{2}+1
\end{equation}%
or $1\leq \left\vert a\right\vert ^{2}\leq 4/3$ that are violated, for
instance,when $a=1/\sqrt{2}.$

\section{Three-player PD game with the GHZ state}

Now, also consider the case when we have the correlation form of players'
payoff relations (\ref{PD_GHZ}) for the considered PD game and for the GHZ
state, we have

\begin{gather}
\xi =P(abc)=1/2,  \notag \\
\lambda =P(a)=1/2,\text{ }\mu =P(b)=1/2,\text{ }\nu =P(c)=1/2,  \notag \\
P(ab)=1,\text{ }P(bc)=1,\text{ }P(ac)=1,\text{ }P(abc)=1/2,
\end{gather}%
which gives us

\begin{equation}
\Pi _{\mathrm{A}}(1/2,1/2,1/2)=\Pi _{\mathrm{B}}(1/2,1/2,1/2)=\Pi _{\mathrm{C%
}}(\lambda ,\mu ,\nu )=3.
\end{equation}

\section{The case of three-qubit pure state}

Consider the state

\begin{gather}
\left\vert \Psi \right\rangle =c_{1}\left\vert 000\right\rangle
+c_{2}\left\vert 001\right\rangle +c_{3}\left\vert 010\right\rangle
+c_{4}\left\vert 011\right\rangle  \notag \\
+c_{5}\left\vert 100\right\rangle +c_{6}\left\vert 101\right\rangle
+c_{7}\left\vert 110\right\rangle +c_{8}\left\vert 111\right\rangle ,
\label{Pure_state}
\end{gather}%
with $\dsum\limits_{i}^{8}\left\vert c_{i}\right\vert ^{2}=1.$ For this
state, we obtain

\begin{eqnarray}
P(ac) &=&\mathrm{Tr}(M_{\mathrm{AC}}^{+1}\mathrm{\rho _{\Psi }})=\mathrm{Tr}%
[\left\vert 0\right\rangle \left\langle 0\right\vert _{\mathrm{A}}\otimes 
\mathcal{I}_{\mathrm{B}}\otimes \left\vert 0\right\rangle \left\langle
0\right\vert _{\mathrm{C}}+\left\vert 1\right\rangle \left\langle
1\right\vert _{\mathrm{A}}\otimes \mathcal{I}_{\mathrm{B}}\otimes \left\vert
1\right\rangle \left\langle 1\right\vert _{\mathrm{C}}]\mathrm{\rho _{\Psi }}
\notag \\
&=&\mathrm{Tr}(\left\vert 000\right\rangle \left\langle 000\right\vert
+\left\vert 010\right\rangle \left\langle 010\right\vert +\left\vert
101\right\rangle \left\langle 101\right\vert +\left\vert 111\right\rangle
\left\langle 111\right\vert )\left\vert \Psi \right\rangle \left\langle \Psi
\right\vert  \notag \\
&=&\left\vert c_{1}\right\vert ^{2}+\left\vert c_{3}\right\vert
^{2}+\left\vert c_{6}\right\vert ^{2}+\left\vert c_{8}\right\vert ^{2}.
\end{eqnarray}%
In analogy with the calculations above, we also obtain

\begin{gather}
\lambda =\left\vert c_{1}\right\vert ^{2}+\left\vert c_{2}\right\vert
^{2}+\left\vert c_{3}\right\vert ^{2}+\left\vert c_{4}\right\vert ^{2}, 
\notag \\
\mu =\left\vert c_{1}\right\vert ^{2}+\left\vert c_{2}\right\vert
^{2}+\left\vert c_{5}\right\vert ^{2}+\left\vert c_{6}\right\vert ^{2}, 
\notag \\
\nu =\left\vert c_{1}\right\vert ^{2}+\left\vert c_{3}\right\vert
^{2}+\left\vert c_{5}\right\vert ^{2}+\left\vert c_{7}\right\vert ^{2}, 
\notag \\
P(ab)=\left\vert c_{1}\right\vert ^{2}+\left\vert c_{2}\right\vert
^{2}+\left\vert c_{7}\right\vert ^{2}+\left\vert c_{8}\right\vert ^{2}, 
\notag \\
P(bc)=\left\vert c_{1}\right\vert ^{2}+\left\vert c_{4}\right\vert
^{2}+\left\vert c_{5}\right\vert ^{2}+\left\vert c_{8}\right\vert ^{2}, 
\notag \\
P(ac)=\left\vert c_{1}\right\vert ^{2}+\left\vert c_{3}\right\vert
^{2}+\left\vert c_{6}\right\vert ^{2}+\left\vert c_{8}\right\vert ^{2}, 
\notag \\
\xi =P(abc)=\left\vert c_{1}\right\vert ^{2}+\left\vert c_{4}\right\vert
^{2}+\left\vert c_{6}\right\vert ^{2}+\left\vert c_{7}\right\vert ^{2}.
\label{Marginals_Coeffs}
\end{gather}

For PD, substituting from Eqs. (\ref{Marginals_Coeffs}) to Eqs. (\ref{PD_GHZ}%
) we then obtain

\begin{gather}
\Pi _{\mathrm{A}}(\lambda ,\mu ,\nu )=2\left[ 3\left\vert c_{1}\right\vert
^{2}+\left\vert c_{2}\right\vert ^{2}+\left\vert c_{3}\right\vert
^{2}+4\left\vert c_{5}\right\vert ^{2}+2\left\vert c_{6}\right\vert
^{2}+2\left\vert c_{7}\right\vert ^{2}-\left\vert c_{8}\right\vert ^{2}%
\right] +1,  \notag \\
\Pi _{\mathrm{B}}(\lambda ,\mu ,\nu )=2\left[ 3\left\vert c_{1}\right\vert
^{2}+\left\vert c_{2}\right\vert ^{2}+4\left\vert c_{3}\right\vert
^{2}+2\left\vert c_{4}\right\vert ^{2}+\left\vert c_{5}\right\vert
^{2}+2\left\vert c_{7}\right\vert ^{2}-\left\vert c_{8}\right\vert ^{2}%
\right] +1,  \notag \\
\Pi _{\mathrm{C}}(\lambda ,\mu ,\nu )=2\left[ 3\left\vert c_{1}\right\vert
^{2}+4\left\vert c_{2}\right\vert ^{2}+\left\vert c_{3}\right\vert
^{2}+2\left\vert c_{4}\right\vert ^{2}+\left\vert c_{5}\right\vert
^{2}+2\left\vert c_{6}\right\vert ^{2}-\left\vert c_{8}\right\vert ^{2}%
\right] +1,  \label{PD_payoffs_coeffs}
\end{gather}%
expressing the player payoffs in terms of the coefficients of the considered
pure state.

\section{The case of mixed three-qubit states}

We consider the mixed state $\mathrm{\rho }_{\mathrm{\Phi }}$ given by

\begin{eqnarray}
\mathrm{\rho }_{\mathrm{\Phi }} &=&p_{1}\left\vert 000\right\rangle
\left\langle 000\right\vert +p_{2}\left\vert 001\right\rangle \left\langle
001\right\vert +p_{3}\left\vert 010\right\rangle \left\langle 010\right\vert
+p_{4}\left\vert 011\right\rangle \left\langle 011\right\vert +  \notag \\
&&p_{5}\left\vert 100\right\rangle \left\langle 100\right\vert
+p_{6}\left\vert 101\right\rangle \left\langle 101\right\vert
+p_{7}\left\vert 110\right\rangle \left\langle 110\right\vert
+p_{8}\left\vert 111\right\rangle \left\langle 111\right\vert ,
\end{eqnarray}%
where $0\leq p_{i}\leq 1$ and $\dsum\limits_{i}^{8}p_{i}=1.$

\begin{eqnarray}
\lambda &=&\mathrm{Tr}(\mathrm{M_{\mathrm{A}}^{0}\rho _{\Phi }})=\mathrm{Tr}%
\left[ (\left\vert 0\right\rangle \left\langle 0\right\vert _{\mathrm{A}%
}\otimes \mathcal{I}_{\mathrm{B}}\otimes \mathcal{I}_{\mathrm{C}})\mathrm{%
\rho _{\Phi }}\right]  \notag \\
&=&\mathrm{Tr}\left[ (\left\vert 000\right\rangle \left\langle
000\right\vert +\left\vert 001\right\rangle \left\langle 001\right\vert
+\left\vert 010\right\rangle \left\langle 010\right\vert +\left\vert
011\right\rangle \left\langle 011\right\vert )\mathrm{\rho _{\mathrm{\Phi }}}%
\right]  \notag \\
&=&p_{1}+p_{2}+p_{3}+p_{4}
\end{eqnarray}

\begin{eqnarray}
\mu &=&\mathrm{Tr}(\mathrm{M_{\mathrm{B}}^{0}\rho _{\Psi }})=\mathrm{Tr}%
\left[ (\mathcal{I}_{\mathrm{A}}\otimes \left\vert 0\right\rangle
\left\langle 0\right\vert _{\mathrm{B}}\otimes \mathcal{I}_{\mathrm{C}})%
\mathrm{\rho }_{\mathrm{\Phi }}\right]  \notag \\
&=&\mathrm{Tr}[p_{1}\left\vert 000\right\rangle \left\langle 000\right\vert
+p_{2}\left\vert 001\right\rangle \left\langle 001\right\vert
+p_{5}\left\vert 100\right\rangle \left\langle 100\right\vert
+p_{6}\left\vert 101\right\rangle \left\langle 101\right\vert ]  \notag \\
&=&p_{1}+p_{2}+p_{5}+p_{6}
\end{eqnarray}

\begin{eqnarray}
\nu &=&\mathrm{Tr}(\mathrm{M_{\mathrm{C}}^{0}\rho _{\Phi }})=\mathrm{Tr}%
\left[ (\mathcal{I}_{\mathrm{A}}\otimes \mathcal{I}_{\mathrm{B}}\otimes
\left\vert 0\right\rangle \left\langle 0\right\vert _{\mathrm{C}})\mathrm{%
\rho _{\Phi }}\right]  \notag \\
&=&\mathrm{Tr}\left[ ((\left\vert 0\right\rangle \left\langle 0\right\vert _{%
\mathrm{A}}+\left\vert 1\right\rangle \left\langle 1\right\vert _{\mathrm{A}%
})\otimes (\left\vert 0\right\rangle \left\langle 0\right\vert _{\mathrm{B}%
}+\left\vert 1\right\rangle \left\langle 1\right\vert _{\mathrm{B}})\otimes
\left\vert 0\right\rangle \left\langle 0\right\vert _{\mathrm{C}})\mathrm{%
\rho _{\Phi }}\right]  \notag \\
&=&p_{1}+p_{3}+p_{5}+p_{7}
\end{eqnarray}

\section{The product state}

Consider a one-qubit pure quantum states

\begin{eqnarray}
\left\vert \psi _{1}\right\rangle &=&e^{i\delta _{1}}(\cos \frac{\theta _{1}%
}{2}\left\vert 0\right\rangle +e^{i\phi _{1}}\sin \frac{\theta _{1}}{2}%
\left\vert 1\right\rangle ),  \notag \\
\left\vert \psi _{2}\right\rangle &=&e^{i\delta _{2}}(\cos \frac{\theta _{2}%
}{2}\left\vert 0\right\rangle +e^{i\phi _{2}}\sin \frac{\theta _{2}}{2}%
\left\vert 1\right\rangle ),  \notag \\
\left\vert \psi _{3}\right\rangle &=&e^{i\delta _{3}}(\cos \frac{\theta _{3}%
}{2}\left\vert 0\right\rangle +e^{i\phi _{3}}\sin \frac{\theta _{3}}{2}%
\left\vert 1\right\rangle ).
\end{eqnarray}%
A product state of these is

\begin{eqnarray}
\left\vert \psi _{1,2,3}\right\rangle &=&e^{i(\delta _{1}+\delta _{2}+\delta
_{3})}[(\cos \frac{\theta _{1}}{2}\cos \frac{\theta _{2}}{2}\cos \frac{%
\theta _{3}}{2}\left\vert 000\right\rangle +e^{i\phi _{3}}\cos \frac{\theta
_{1}}{2}\cos \frac{\theta _{2}}{2}\sin \frac{\theta _{3}}{2}\left\vert
001\right\rangle +  \notag \\
&&e^{i\phi _{2}}\cos \frac{\theta _{1}}{2}\sin \frac{\theta _{2}}{2}\cos 
\frac{\theta _{3}}{2}\left\vert 010\right\rangle +e^{i(\phi _{2}+\phi
_{3})}\cos \frac{\theta _{1}}{2}\sin \frac{\theta _{2}}{2}\sin \frac{\theta
_{3}}{2}\left\vert 011\right\rangle )+  \notag \\
&&(e^{i\phi _{1}}\sin \frac{\theta _{1}}{2}\cos \frac{\theta _{2}}{2}\cos 
\frac{\theta _{3}}{2}\left\vert 100\right\rangle +e^{i(\phi _{1}+\phi
_{3})}\sin \frac{\theta _{1}}{2}\cos \frac{\theta _{2}}{2}\sin \frac{\theta
_{3}}{2}\left\vert 101\right\rangle +  \notag \\
&&e^{i(\phi _{1}+\phi _{2})}\sin \frac{\theta _{1}}{2}\sin \frac{\theta _{2}%
}{2}\cos \frac{\theta _{3}}{2}\left\vert 110\right\rangle +e^{i(\phi
_{2}+\phi _{3})}e^{i\phi _{1}}\sin \frac{\theta _{1}}{2}\sin \frac{\theta
_{2}}{2}\sin \frac{\theta _{3}}{2}\left\vert 111\right\rangle )],  \notag \\
&&  \label{Product_State}
\end{eqnarray}%
and for the pure state $\left\vert \Psi \right\rangle =c_{1}\left\vert
000\right\rangle +c_{2}\left\vert 001\right\rangle +c_{3}\left\vert
010\right\rangle +c_{4}\left\vert 011\right\rangle +c_{5}\left\vert
100\right\rangle +c_{6}\left\vert 101\right\rangle +c_{7}\left\vert
110\right\rangle +c_{8}\left\vert 111\right\rangle $ we have $\lambda
=\left\vert c_{1}\right\vert ^{2}+\left\vert c_{2}\right\vert
^{2}+\left\vert c_{3}\right\vert ^{2}+\left\vert c_{4}\right\vert ^{2}$. For
the product state (\ref{Product_State}) we therefore obtain

\begin{equation}
\lambda =\cos ^{2}\frac{\theta _{1}}{2},\text{ }\mu =\cos ^{2}\frac{\theta
_{2}}{2},\text{ }\nu =\cos ^{2}\frac{\theta _{3}}{2}.
\end{equation}%
Similarly,

\begin{eqnarray}
P(ab) &=&\left\vert c_{1}\right\vert ^{2}+\left\vert c_{2}\right\vert
^{2}+\left\vert c_{7}\right\vert ^{2}+\left\vert c_{8}\right\vert ^{2} 
\notag \\
&=&\cos ^{2}\frac{\theta _{1}}{2}\cos ^{2}\frac{\theta _{2}}{2}+\sin ^{2}%
\frac{\theta _{1}}{2}\sin ^{2}\frac{\theta _{2}}{2},
\end{eqnarray}%
and that can be expressed as $P(ab)=\lambda \mu +(1-\lambda )(1-\mu ).$ Also,

\begin{eqnarray}
P(bc) &=&\left\vert c_{1}\right\vert ^{2}+\left\vert c_{4}\right\vert
^{2}+\left\vert c_{5}\right\vert ^{2}+\left\vert c_{8}\right\vert ^{2} 
\notag \\
&=&\cos ^{2}\frac{\theta _{2}}{2}\cos ^{2}\frac{\theta _{3}}{2}+\sin ^{2}%
\frac{\theta _{2}}{2}\sin ^{2}\frac{\theta _{3}}{2},
\end{eqnarray}%
and for the state (\ref{Product_State}) this can be expressed as $P(bc)=\mu
\nu +(1-\mu )(1-\nu ).$ Similarly,

\begin{eqnarray}
P(ac) &=&\left\vert c_{1}\right\vert ^{2}+\left\vert c_{3}\right\vert
^{2}+\left\vert c_{6}\right\vert ^{2}+\left\vert c_{8}\right\vert ^{2} 
\notag \\
&=&\cos ^{2}\frac{\theta _{1}}{2}\cos ^{2}\frac{\theta _{3}}{2}+\sin ^{2}%
\frac{\theta _{1}}{2}\sin ^{2}\frac{\theta _{3}}{2},
\end{eqnarray}%
that can be expressed as $P(ac)=\lambda \nu +(1-\lambda )(1-\nu ),$ and

\begin{eqnarray}
P(abc) &=&\left\vert c_{1}\right\vert ^{2}+\left\vert c_{4}\right\vert
^{2}+\left\vert c_{6}\right\vert ^{2}+\left\vert c_{7}\right\vert ^{2} 
\notag \\
&=&\lambda \mu \nu +\lambda (1-\mu )(1-\nu )+(1-\lambda )\mu (1-\nu
)+(1-\lambda )(1-\mu )\nu ,
\end{eqnarray}%
for the state (\ref{Product_State}).

\section{Bell's inequalities violation}

The above case shows that a set of non-factorizable probabilities,
understood as above, can emerge from a factorizable quantum state of three
qubits. This is because $\lambda =P(a),$ $\mu =P(b),$ $\nu =P(c)$ and for
the product state (\ref{Product_State}) we find, for instance, $P(abc)\neq
P(a)P(b)P(c)$. This set, however, should satisfy Bell's inequalities. Below,
we show that this is the case. Considering BIs (\ref{BIs}), we note that for
the state (\ref{Product_State}) BIs take the form

\begin{gather}
3\left[ \lambda +\mu +\nu \right] \leq 4+2\left[ \lambda \mu +\mu \nu
+\lambda \nu \right] ,  \notag \\
1+2\lambda (\mu +\nu )\leq 3\lambda +2\mu \nu ,  \notag \\
1+2\mu (\lambda +\nu )\leq 3\mu +2\lambda \nu ,  \notag \\
1+2\nu (\lambda +\mu )\leq 3\nu +2\lambda \mu .  \label{BIs_Product_State}
\end{gather}%
The last three inequalities added give us

\begin{gather}
3+2[\lambda (\mu +\nu )+\mu (\lambda +\nu )+\nu (\lambda +\mu )]\leq
3(\lambda +\mu +\nu )+2(\lambda \mu +\lambda \nu +\mu \nu ),  \notag \\
3+4[\lambda \mu +\lambda \nu +\mu \nu ]\leq 3(\lambda +\mu +\nu )+2(\lambda
\mu +\lambda \nu +\mu \nu ),  \notag \\
3+2[\lambda \mu +\lambda \nu +\mu \nu ]\leq 3(\lambda +\mu +\nu ),  \notag \\
4+2[\lambda \mu +\lambda \nu +\mu \nu ]\leq 1+3(\lambda +\mu +\nu ),
\end{gather}%
the last inequality above combined with the last inequality in the previous
set give

\begin{equation}
3\left[ \lambda +\mu +\nu \right] \leq 4+2\left[ \lambda \mu +\mu \nu
+\lambda \nu \right] \leq 1+3(\lambda +\mu +\nu ),
\end{equation}%
that is, the set of inequalities are self consistent. In contrast, for an
entangled state, a similar set of BIs would not be self-consistent i.e.
holding of three inequalities would lead to contradicting with the fourth
one. Note that even for an entangled state, Bell's inequalities are violated
only in some directions and not in every direction.

In contrast, consider the GHZ state for which we have

\begin{gather}
\xi =P(abc)=1/2,  \notag \\
\lambda =P(a)=1/2,\text{ }\mu =P(b)=1/2,\text{ }\nu =P(c)=1/2,  \notag \\
P(ab)=1,\text{ }P(bc)=1,\text{ }P(ac)=1,\text{ }P(abc)=1/2.
\end{gather}%
and the BIs

\begin{align}
P(a)+P(b)+P(c)& \leq 1+P(ab)+P(ac)+P(bc),  \notag \\
P(ab)+P(ac)& \leq P(a)+P(bc),  \notag \\
P(ab)+P(bc)& \leq P(b)+P(ac),  \notag \\
P(ac)+P(bc)& \leq P(c)+P(ab),
\end{align}%
take the form $3/2\leq 4,$ $2\leq 3/2,$ $2\leq 3/2,$ $2\leq 3/2,3/2\leq 4,$ $%
2\leq 3/2,$ $2\leq 3/2,$ $2\leq 3/2,$ and BIs are, therefore, violated.

\section{Quantum solution of three-player PD}

We consider the mixed-strategy players' payoffs in the three-player PD as
given by (\ref{PD_GHZ})

\begin{gather}
\Pi _{\mathrm{A}}(\lambda ,\mu ,\nu )=\xi -P(ab)-P(ac)-\lambda +4\mu +4\nu
+1,  \notag \\
\Pi _{\mathrm{B}}(\lambda ,\mu ,\nu )=\xi -P(ab)-P(bc)+4\lambda -\mu +4\nu
+1,  \notag \\
\Pi _{\mathrm{C}}(\lambda ,\mu ,\nu )=\xi -P(bc)-P(ac)+4\lambda +4\mu -\nu
+1,  \label{payoffs_mixed_classical}
\end{gather}%
and analyze these for various quantum states.

\subsection{The factorizable state}

For the factorizable state (\ref{Product_State}), we have

\begin{gather}
P(a)=\lambda ,\text{ }P(b)=\mu ,\text{ }P(c)=\nu ,\text{ }P(ab)=\lambda \mu
+(1-\lambda )(1-\mu ),  \notag \\
P(bc)=\mu \nu +(1-\mu )(1-\nu ),\text{ }P(ac)=\lambda \nu +(1-\lambda
)(1-\nu ),  \notag \\
P(abc)=\lambda \mu \nu +\lambda (1-\mu )(1-\nu )+(1-\lambda )\mu (1-\nu
)+(1-\lambda )(1-\mu )\nu ,
\end{gather}%
with which the payoffs (\ref{payoffs_mixed_classical}) can be re-expressed as

\begin{gather}
\Pi _{\mathrm{A}}(\lambda ,\mu ,\nu )=2(2\lambda \mu \nu -2\lambda \mu
-2\lambda \nu -\mu \nu +\lambda +3\mu +3\nu )-1,  \notag \\
\Pi _{\mathrm{B}}(\lambda ,\mu ,\nu )=2(2\lambda \mu \nu -2\lambda \mu
-\lambda \nu -2\mu \nu +3\lambda +\mu +3\nu )-1,  \notag \\
\Pi _{\mathrm{C}}(\lambda ,\mu ,\nu )=2(2\lambda \mu \nu -\lambda \mu
-2\lambda \nu -2\mu \nu +3\lambda +3\mu +\nu )-1,
\end{gather}%
and Nash inequalities then take the form

\begin{eqnarray}
\frac{\partial \Pi _{\mathrm{A}}}{\partial \lambda } &\mid &_{(\lambda
^{\ast },\mu ^{\ast },\nu ^{\ast })}(\lambda ^{\ast }-\lambda )=2(2\mu
^{\ast }\nu ^{\ast }-2\mu ^{\ast }-2\nu ^{\ast }+1)(\lambda ^{\ast }-\lambda
)\geq 0,  \notag \\
\frac{\partial \Pi _{\mathrm{B}}}{\partial \mu } &\mid &_{(\lambda ^{\ast
},\mu ^{\ast },\nu ^{\ast })}(\mu ^{\ast }-\mu )=2(2\lambda ^{\ast }\nu
^{\ast }-2\lambda ^{\ast }-2\nu ^{\ast }+1)(\mu ^{\ast }-\mu )\geq 0,  \notag
\\
\frac{\partial \Pi _{\mathrm{C}}}{\partial \nu } &\mid &_{(\lambda ^{\ast
},\mu ^{\ast },\nu ^{\ast })}(\nu ^{\ast }-\nu )=2(2\lambda ^{\ast }\mu
^{\ast }-2\lambda ^{\ast }-2\mu ^{\ast }+1)(\nu ^{\ast }-\nu )\geq 0.
\end{eqnarray}%
that can be expressed as

\begin{eqnarray}
(\mu ^{\ast }\nu ^{\ast }-\mu ^{\ast }-\nu ^{\ast }+1/2)(\lambda ^{\ast
}-\lambda ) &\geq &0,  \notag \\
(\lambda ^{\ast }\nu ^{\ast }-\lambda ^{\ast }-\nu ^{\ast }+1/2)(\mu ^{\ast
}-\mu ) &\geq &0,  \notag \\
(\lambda ^{\ast }\mu ^{\ast }-\lambda ^{\ast }-\mu ^{\ast }+1/2)(\nu ^{\ast
}-\nu ) &\geq &0.  \label{NIs_product state}
\end{eqnarray}%
From (\ref{NIs_product state}) neither $(\lambda ^{\ast },\mu ^{\ast },\nu
^{\ast })=(1,1,1)$ nor $(\lambda ^{\ast },\mu ^{\ast },\nu ^{\ast })=(0,0,0)$
emerge as a solution. For the case of non-edge triples $(\lambda ^{\ast
},\mu ^{\ast },\nu ^{\ast })$, we have

\begin{eqnarray}
\mu ^{\ast }\nu ^{\ast }-\mu ^{\ast }-\nu ^{\ast }+1/2 &=&0,  \label{1} \\
\lambda ^{\ast }\nu ^{\ast }-\lambda ^{\ast }-\nu ^{\ast }+1/2 &=&0,
\label{2} \\
\lambda ^{\ast }\mu ^{\ast }-\lambda ^{\ast }-\mu ^{\ast }+1/2 &=&0.
\label{3}
\end{eqnarray}%
Now, subtracting (\ref{2}) from (\ref{1}), subtracting (\ref{3}) from (\ref%
{2}), and subtracting (\ref{3}) from (\ref{1}) gives

\begin{equation}
(\nu ^{\ast }-1)(\mu ^{\ast }-\lambda ^{\ast })=0,\text{ }(\lambda ^{\ast
}-1)(\nu ^{\ast }-\mu ^{\ast })=0,\text{ }(\mu ^{\ast }-1)(\nu ^{\ast
}-\lambda ^{\ast })=0,  \label{4-1}
\end{equation}%
and it follows from (\ref{4-1}) that $\lambda ^{\ast }=\mu ^{\ast }=\nu
^{\ast }$ holds for the solution. Also, Eqs (\ref{1},\ref{2},\ref{3}) can be
expressed as

\begin{equation}
\mu ^{\ast }+\nu ^{\ast }-\mu ^{\ast }\nu ^{\ast }=1/2,\text{ }\lambda
^{\ast }+\nu ^{\ast }-\lambda ^{\ast }\nu ^{\ast }=1/2,\text{ }\lambda
^{\ast }+\mu ^{\ast }-\lambda ^{\ast }\mu ^{\ast }=1/2.
\end{equation}%
From $\lambda ^{\ast }=\mu ^{\ast }=\nu ^{\ast }$, we obtain $2\lambda
^{\ast }-(\lambda ^{\ast })^{2}=1/2$ or $(\lambda ^{\ast })^{2}-2\lambda
^{\ast }+1/2=0$ or $\lambda ^{\ast }=\frac{2\pm \sqrt{2}}{2}$ and as $0\leq
\lambda ^{\ast },\mu ^{\ast },\nu ^{\ast }\leq 1$, we have $\lambda ^{\ast
}=\mu ^{\ast }=\nu ^{\ast }=\frac{2-\sqrt{2}}{2}$ and then

\begin{eqnarray}
P(ab) &=&\lambda ^{\ast }\mu ^{\ast }+(1-\lambda ^{\ast })(1-\mu ^{\ast })=2-%
\sqrt{2},  \notag \\
P(bc) &=&\mu ^{\ast }\nu ^{\ast }+(1-\mu ^{\ast })(1-\nu ^{\ast })=2-\sqrt{2}%
,  \notag \\
P(ac) &=&\lambda ^{\ast }\nu ^{\ast }+(1-\lambda ^{\ast })(1-\nu ^{\ast })=2-%
\sqrt{2},  \notag \\
P(abc) &=&\lambda ^{\ast }[\mu ^{\ast }\nu ^{\ast }+(1-\mu ^{\ast })(1-\nu
^{\ast })]+\mu ^{\ast }(1-\lambda ^{\ast })(1-\nu ^{\ast })+\nu ^{\ast
}(1-\lambda ^{\ast })(1-\mu ^{\ast }),  \notag \\
&=&(2-\sqrt{2})(3-\sqrt{2})/2.
\end{eqnarray}%
To find the quantum states corresponding to these marginals, we use the
state (\ref{Pure_state}) for which, the mixed-strategy solution $\lambda
^{\ast }=\mu ^{\ast }=\nu ^{\ast }=\frac{2-\sqrt{2}}{2}$ results in negative
value for the probability $p_{4}$. This solution is to be discarded and the
factorizable state (\ref{Product_State}) therefore generates no solution. We
now consider GHZ and W-type states.

\subsection{GHZ-type state}

For GHZ type state i.e. $\left\vert \Psi \right\rangle =a\left\vert
000\right\rangle +b\left\vert 111\right\rangle $ where $a,b\in $ \QTR{cal}{%
\textrm{C}} and $\left\vert a\right\vert ^{2}+\left\vert b\right\vert ^{2}=1$%
. This gives $\left\langle \Psi \right\vert =a^{\ast }\left\langle
000\right\vert +b^{\ast }\left\langle 111\right\vert $ and we have

\begin{equation}
\rho _{\Psi }=\left\vert \Psi \right\rangle \left\langle \Psi \right\vert
=\left\vert a\right\vert ^{2}\left\vert 000\right\rangle \left\langle
000\right\vert +ab^{\ast }\left\vert 000\right\rangle \left\langle
111\right\vert +a^{\ast }b\left\vert 111\right\rangle \left\langle
000\right\vert +\left\vert b\right\vert ^{2}\left\vert 111\right\rangle
\left\langle 111\right\vert ,
\end{equation}%
and from (\ref{strategies_as_traces}) we as considered above, therefore, have

\begin{equation}
\lambda =P(a)=\mathrm{Tr}(M_{\mathrm{A}}^{0}\rho ),\mu =P(b)=\mathrm{Tr}(M_{%
\mathrm{B}}^{0}\rho ),\nu =P(c)=\mathrm{Tr}(M_{\mathrm{C}}^{0}\rho ).
\end{equation}

\QTP{Body Math}
\begin{eqnarray}
\lambda &=&\mathrm{Tr}(\mathrm{M_{\mathrm{A}}^{0}\rho _{\Psi }})=\mathrm{Tr}%
\left[ (\left\vert 0\right\rangle \left\langle 0\right\vert _{\mathrm{A}%
}\otimes \mathcal{I}_{\mathrm{B}}\otimes \mathcal{I}_{\mathrm{C}})\mathrm{%
\rho _{\Psi }}\right] =\left\vert a\right\vert ^{2},  \notag \\
\mu &=&\mathrm{Tr}(\mathrm{M_{\mathrm{B}}^{0}\rho _{\Psi }})=\mathrm{Tr}%
\left[ (\mathcal{I}_{\mathrm{A}}\otimes \left\vert 0\right\rangle
\left\langle 0\right\vert _{\mathrm{B}}\otimes \mathcal{I}_{\mathrm{C}})%
\mathrm{\rho _{\Psi }}\right] =\left\vert a\right\vert ^{2},  \notag \\
\nu &=&\mathrm{Tr}(\mathrm{M_{\mathrm{C}}^{0}\rho _{\Psi }})=\mathrm{Tr}%
\left[ (\mathcal{I}_{\mathrm{A}}\otimes \mathcal{I}_{\mathrm{B}}\otimes
\left\vert 0\right\rangle \left\langle 0\right\vert _{\mathrm{C}})\mathrm{%
\rho _{\Psi }}\right] =\left\vert a\right\vert ^{2},
\end{eqnarray}

\begin{eqnarray}
P(ab) &=&\mathrm{Tr}(M_{\mathrm{AB}}^{+1}\mathrm{\rho _{\Psi }})=\mathrm{Tr}%
[(\left\vert 0\right\rangle \left\langle 0\right\vert _{\mathrm{A}}\otimes
\left\vert 0\right\rangle \left\langle 0\right\vert _{\mathrm{B}}+\left\vert
1\right\rangle \left\langle 1\right\vert _{\mathrm{A}}\otimes \left\vert
1\right\rangle \left\langle 1\right\vert _{\mathrm{B}})\otimes \mathcal{I}_{%
\mathrm{C}}]\mathrm{\rho _{\Psi }}=1,  \notag \\
P(bc) &=&\mathrm{Tr}(M_{\mathrm{BC}}^{+1}\mathrm{\rho _{\Psi }})=\mathrm{Tr}[%
\mathcal{I}_{\mathrm{A}}\otimes (\left\vert 0\right\rangle \left\langle
0\right\vert _{\mathrm{B}}\otimes \left\vert 0\right\rangle \left\langle
0\right\vert _{\mathrm{C}}+\left\vert 1\right\rangle \left\langle
1\right\vert _{\mathrm{B}}\otimes \left\vert 1\right\rangle \left\langle
1\right\vert _{\mathrm{C}})]\mathrm{\rho _{\Psi }}=1,  \notag \\
P(ac) &=&\mathrm{Tr}(M_{\mathrm{AC}}^{+1}\mathrm{\rho _{\Psi }})=\mathrm{Tr}%
[\left\vert 0\right\rangle \left\langle 0\right\vert _{\mathrm{A}}\otimes 
\mathcal{I}_{\mathrm{B}}\otimes \left\vert 0\right\rangle \left\langle
0\right\vert _{\mathrm{C}}+\left\vert 1\right\rangle \left\langle
1\right\vert _{\mathrm{A}}\otimes \mathcal{I}_{\mathrm{B}}\otimes \left\vert
1\right\rangle \left\langle 1\right\vert _{\mathrm{C}}]\mathrm{\rho _{\Psi }}%
=1,  \notag \\
\xi &=&P(abc)=\mathrm{Tr}(M_{\mathrm{ABC}}^{+1}\mathrm{\rho _{\Psi }}%
)=\left\vert a\right\vert ^{2}.
\end{eqnarray}%
As the above probabilities are over determined, the GHZ state does not
permit players making choices within the considered scheme.

\subsection{W-type state}

For the W-type state $\left\vert \Psi _{W}\right\rangle =c_{2}\left\vert
001\right\rangle +c_{3}\left\vert 010\right\rangle +c_{5}\left\vert
100\right\rangle ,$ with $\left\vert c_{2}\right\vert ^{2}+\left\vert
c_{3}\right\vert ^{2}+\left\vert c_{5}\right\vert ^{2}=1,$ we obtain

\begin{equation}
\lambda =\left\vert c_{2}\right\vert ^{2}+\left\vert c_{3}\right\vert ^{2},%
\text{ }\mu =\left\vert c_{2}\right\vert ^{2}+\left\vert c_{5}\right\vert
^{2},\text{ }\nu =\left\vert c_{3}\right\vert ^{2}+\left\vert
c_{5}\right\vert ^{2},
\end{equation}%
and

\begin{equation}
P(ab)=\left\vert c_{2}\right\vert ^{2},\text{ }P(bc)=\left\vert
c_{5}\right\vert ^{2},\text{ }P(ac)=\left\vert c_{3}\right\vert ^{2},\text{ }%
\xi =P(abc)=0.
\end{equation}%
We note that

\begin{gather}
\lambda +\mu +\nu =2(\left\vert c_{2}\right\vert ^{2}+\left\vert
c_{3}\right\vert ^{2}+\left\vert c_{5}\right\vert ^{2})=2,
\label{w_type_first} \\
P(ab)=\frac{1}{2}(\lambda -\nu +\mu ),\text{ }P(bc)=\frac{1}{2}(\mu +\nu
-\lambda ),\text{ }P(ac)=\frac{1}{2}(\lambda +\nu -\mu ),
\end{gather}%
and the mixed-strategy players' payoffs (\ref{PD_GHZ}) in three-player PD
take the form

\begin{eqnarray}
\Pi _{\mathrm{A}}(\lambda ,\mu ,\nu ) &=&-2\lambda +4\mu +4\nu +1,  \notag \\
\Pi _{\mathrm{B}}(\lambda ,\mu ,\nu ) &=&4\lambda -2\mu +4\nu +1,  \notag \\
\Pi _{\mathrm{C}}(\lambda ,\mu ,\nu ) &=&4\lambda +4\mu -2\nu +1.
\label{Payoffs_wState}
\end{eqnarray}%
Nash inequalities then become

\begin{eqnarray}
\frac{\partial \Pi _{\mathrm{A}}}{\partial \lambda } &\mid &_{(\lambda
^{\ast },\mu ^{\ast },\nu ^{\ast })}(\lambda ^{\ast }-\lambda )=-2(\lambda
^{\ast }-\lambda )\geq 0,  \notag \\
\frac{\partial \Pi _{\mathrm{B}}}{\partial \mu } &\mid &_{(\lambda ^{\ast
},\mu ^{\ast },\nu ^{\ast })}(\mu ^{\ast }-\mu )=-2(\mu ^{\ast }-\mu )\geq 0,
\notag \\
\frac{\partial \Pi _{\mathrm{C}}}{\partial \nu } &\mid &_{(\lambda ^{\ast
},\mu ^{\ast },\nu ^{\ast })}(\nu ^{\ast }-\nu )=-2(\nu ^{\ast }-\nu )\geq 0.
\end{eqnarray}%
giving the outcome as $(\lambda ^{\ast },\mu ^{\ast },\nu ^{\ast })=(0,0,0)$
which contradicts Eq. (\ref{w_type_first}). The W-type state, therefore,
does not offer a resolution or representation of three-player PD.

\subsection{The state $\left\vert \Psi \right\rangle =c_{4}\left\vert
011\right\rangle +c_{6}\left\vert 101\right\rangle +c_{7}\left\vert
110\right\rangle $}

For the state

\begin{gather}
\left\vert \Psi \right\rangle =c_{4}\left\vert 011\right\rangle
+c_{6}\left\vert 101\right\rangle +c_{7}\left\vert 110\right\rangle ,  \notag
\\
\left\vert c_{4}\right\vert ^{2}+\left\vert c_{6}\right\vert ^{2}+\left\vert
c_{7}\right\vert ^{2}=1,  \label{PD-state}
\end{gather}

we obtain from Eqs. (\ref{Marginals_Coeffs}, \ref{PD_payoffs_coeffs})

\begin{gather}
\lambda =\left\vert c_{4}\right\vert ^{2},\text{ }\mu =\left\vert
c_{6}\right\vert ^{2},\text{ }\nu =\left\vert c_{7}\right\vert ^{2},  \notag
\\
\Pi _{\mathrm{A}}(\lambda ,\mu ,\nu )=4\left( \mu +\nu \right) +1,  \notag \\
\Pi _{\mathrm{B}}(\lambda ,\mu ,\nu )=4\left( \lambda +\nu \right) +1, 
\notag \\
\Pi _{\mathrm{C}}(\lambda ,\mu ,\nu )=4\left( \lambda +\mu \right) +1,
\label{Payoffs_theState}
\end{gather}%
and thus

\begin{equation}
\frac{\partial \Pi _{\mathrm{A}}}{\partial \lambda }\mid _{(\lambda ^{\ast
},\mu ^{\ast },\nu ^{\ast })}=\frac{\partial \Pi _{\mathrm{B}}}{\partial \mu 
}\mid _{(\lambda ^{\ast },\mu ^{\ast },\nu ^{\ast })}=\frac{\partial \Pi _{%
\mathrm{C}}}{\partial \nu }\mid _{(\lambda ^{\ast },\mu ^{\ast },\nu ^{\ast
})}=0,
\end{equation}%
resulting in the strategy triple $(\lambda ^{\ast },\mu ^{\ast },\nu ^{\ast
})$ becoming a NE, with the players' payoffs obtained from Eq. (\ref%
{Payoffs_theState}) as

\begin{gather}
\Pi _{\mathrm{A}}(\lambda ^{\ast },\mu ^{\ast },\nu ^{\ast })=4\left( \mu
^{\ast }+\nu ^{\ast }\right) +1,  \notag \\
\Pi _{\mathrm{B}}(\lambda ^{\ast },\mu ^{\ast },\nu ^{\ast })=4\left(
\lambda ^{\ast }+\nu ^{\ast }\right) +1,  \notag \\
\Pi _{\mathrm{C}}(\lambda ^{\ast },\mu ^{\ast },\nu ^{\ast })=4\left(
\lambda ^{\ast }+\mu ^{\ast }\right) +1,  \notag \\
\text{with }\lambda ^{\ast }+\mu ^{\ast }+\nu ^{\ast }=1.
\end{gather}%
That is, this state allows surpassing the strong domination of pure strategy
triple $(\lambda ^{\ast },\mu ^{\ast },\nu ^{\ast })=(0,0,0)$ in
three-player PD but with the unattractive outcome that no unique but a
continuum of NE that emerge. For instance, at $(\lambda ^{\ast },\mu ^{\ast
},\nu ^{\ast })=(1/3,1/3,1/3)$ with $\Pi _{\mathrm{A,B,C}%
}(1/3,1/3,1/3)=11/3. $

\section{The case of a three-player symmetric cooperative game}

Note that $a,b,c$ are bivalent observable, each mapping into $\left\{
+1,-1\right\} .$ Let $\wp $ be an arbitrary subset of $\Re $. The players in 
$\wp $ may form a coalition so that, for all practical purposes, the
coalition $\wp $ appears as a single player. It is expected that players in $%
(\Re -\wp )$ will form an opposing coalition and the game has two opposing
coalition players\ i.e. $\wp $ and $(\Re -\wp )$. Each of the three players 
\textrm{A}, \textrm{B}, and \textrm{C} chooses one of the two strategies $+1$%
, $-1$. If the three players choose the same strategy there is no payoff;
otherwise, the two players who have chosen the same strategy receive one
unit of money each from the `odd man.' Payoff functions $\Pi _{\mathrm{A}}$, 
$\Pi _{\mathrm{B}}$ and $\Pi _{\mathrm{C}}$ for players \textrm{A}, \textrm{B%
} and \textrm{C}, respectively, are given as \cite{BurgerFreund}

\begin{align}
\Pi _{\mathrm{A}}(+1,+1,+1)& =\Pi _{\mathrm{A}}(-1,-1,-1)=0,  \notag \\
\Pi _{\mathrm{A}}(+1,+1,-1)& =\Pi _{\mathrm{A}}(-1,-1,+1)=\Pi _{\mathrm{A}%
}(+1,-1,+1)=\Pi _{\mathrm{A}}(-1,+1,-1)=1,  \notag \\
\Pi _{\mathrm{A}}(+1,-1,-1)& =\Pi _{\mathrm{A}}(-1,+1,+1)=-2,
\label{PayoffsCoop}
\end{align}%
with similar expressions for $\Pi _{\mathrm{B}}$ and $\Pi _{\mathrm{C}}$.

Suppose $\wp =\left\{ \mathrm{B},\mathrm{C}\right\} $, hence $\Re -\wp
=\left\{ \mathrm{A}\right\} $. The coalition game represented by $\Gamma
_{\wp }$ is given by the payoff matrix

\begin{equation}
\wp 
\begin{array}{c}
\left[ +1,+1\right] \\ 
\left[ +1,-1\right] \\ 
\left[ -1,+1\right] \\ 
\left[ -1,-1\right]%
\end{array}%
\overset{\overset{{\LARGE \Re -\wp }}{%
\begin{array}{cc}
\left[ +1\right] & \left[ -1\right]%
\end{array}%
}}{\left( 
\begin{array}{cc}
0 & 2 \\ 
-1 & -1 \\ 
-1 & -1 \\ 
2 & 0%
\end{array}%
\right) .}
\end{equation}%
Here the strategies $\left[ +1,-1\right] $ and $\left[ -1,+1\right] $ are
dominated by $\left[ +1,+1\right] $ and $\left[ -1,-1\right] $. After
eliminating these dominated strategies the payoff matrix becomes

\begin{equation}
\wp 
\begin{array}{c}
\left[ +1,+1\right] \\ 
\left[ -1,-1\right]%
\end{array}%
\overset{\overset{{\LARGE \Re -\wp }}{%
\begin{array}{cc}
\left[ +1\right] & \left[ -1\right]%
\end{array}%
}}{\left( 
\begin{array}{cc}
0 & 2 \\ 
2 & 0%
\end{array}%
\right) }.
\end{equation}%
It is seen that the mixed strategies:

\begin{align}
\wp :& \frac{1}{2}\left[ +1,+1\right] +\frac{1}{2}\left[ -1,-1\right] ,
\label{cltCoop} \\
\Re -\wp :& \frac{1}{2}\left[ +1\right] +\frac{1}{2}\left[ -1\right] \text{.}
\label{lftCoop}
\end{align}%
are optimal for $\wp $ and $(\Re -\wp )$ respectively. In coalition games
the notion of a strategy disappears; the main features are those of a
coalition and the value or worth of the coalition. The underlying assumption
is that each coalition can guarantee its members a certain amount called the 
\emph{value of a coalition}\textit{\ }which measures the worth the coalition
possesses. It is characterized as the payoff which the coalition can assure
for itself by selecting an appropriate strategy, whereas the `odd man' can
prevent the coalition from getting more than this amount. With the
strategies (\ref{cltCoop},\ref{lftCoop}) a payoff $1$ for players $\wp $ is
assured for all strategies of the opponent; hence, the value of the
coalition $\upsilon (\Gamma _{\wp })$ is $1$ i.e. $\upsilon (\left\{ \mathrm{%
B},\mathrm{C}\right\} )=1$. Since $\Gamma $ is a zero-sum game, $\upsilon
(\Gamma _{\wp })$ can also be used to find $\upsilon (\Gamma _{\Re -\wp })$
as $\upsilon (\left\{ \mathrm{A}\right\} )=-1$. The game is symmetric and
one can write

\begin{align}
\upsilon (\Gamma _{\wp })& =1\text{, \ \ and\ \ \ }\upsilon (\Gamma _{\Re
-\wp })=-1,\text{ or}  \notag \\
\upsilon (\left\{ \mathrm{A}\right\} )& =\upsilon (\left\{ \mathrm{B}%
\right\} )=\upsilon (\left\{ \mathrm{C}\right\} )=-1,  \notag \\
\upsilon (\left\{ \mathrm{A},\mathrm{B}\right\} )& =\upsilon (\left\{ 
\mathrm{B},\mathrm{C}\right\} )=\upsilon (\left\{ \mathrm{C},\mathrm{A}%
\right\} )=1.  \label{VcltC}
\end{align}%
The mixed-strategy payoff relations for the players \textrm{A}$\mathrm{,}$%
\textrm{B}$\mathrm{,}$ and \textrm{C} then are defined as

\begin{gather}
\Pi _{\mathrm{A,B,C}}(\lambda ,\mu ,\nu )=\Pi _{\mathrm{A,B,C}%
}(+1,+1,+1)P(abc)+\Pi _{\mathrm{A,B,C}}(+1,+1,-1)P(ab\bar{c})+  \notag \\
\Pi _{\mathrm{A,B,C}}(+1,-1,+1)P(a\bar{b}c)+\Pi _{\mathrm{A,B,C}%
}(+1,-1,-1)P(a\bar{b}\bar{c})+  \notag \\
\Pi _{\mathrm{A,B,C}}(-1,+1,+1)P(\bar{a}bc)+\Pi _{\mathrm{A,B,C}}(-1,+1,-1)P(%
\bar{a}b\bar{c})+  \notag \\
\Pi _{\mathrm{A,B,C}}(-1,-1,+1)P(\bar{a}\bar{b}c)+\Pi _{\mathrm{A,B,C}%
}(-1,-1,-1)P(\bar{a}\bar{b}\bar{c}),
\end{gather}%
where $\lambda ,\mu ,\nu \in \lbrack 0,1]$ and $P(\bullet )$ is the
probability that each enclosed observable takes the value $+1.$ In view of (%
\ref{PayoffsCoop}), the above payoffs can be written as%
\begin{eqnarray}
\Pi _{\mathrm{A}}(\lambda ,\mu ,\nu ) &=&P(ab\bar{c})+P(a\bar{b}c)-2P(a\bar{b%
}\bar{c})-2P(\bar{a}bc)+P(\bar{a}b\bar{c})+P(\bar{a}\bar{b}c),  \notag \\
\Pi _{\mathrm{B}}(\lambda ,\mu ,\nu ) &=&P(ab\bar{c})-2P(a\bar{b}c)+P(a\bar{b%
}\bar{c})+P(\bar{a}bc)-2P(\bar{a}b\bar{c})+P(\bar{a}\bar{b}c),  \notag \\
\Pi _{\mathrm{C}}(\lambda ,\mu ,\nu ) &=&-2P(ab\bar{c})+P(a\bar{b}c)+P(a\bar{%
b}\bar{c})+P(\bar{a}bc)+P(\bar{a}b\bar{c})-2P(\bar{a}\bar{b}c).
\label{mixed_strategy_payoffs}
\end{eqnarray}

\subsection{Factorizable case}

For factorizable probabilities, the payoffs (\ref{mixed_strategy_payoffs})
take the form

\begin{eqnarray}
\Pi _{\mathrm{A}}(\lambda ,\mu ,\nu ) &=&2\lambda (\mu +\nu -1)+\mu +\nu
-4\mu \nu ,  \notag \\
\Pi _{\mathrm{B}}(\lambda ,\mu ,\nu ) &=&2\mu (\nu +\lambda -1)+\lambda +\nu
-4\lambda \nu ,  \notag \\
\Pi _{\mathrm{C}}(\lambda ,\mu ,\nu ) &=&2\nu (\lambda +\mu -1)+\lambda +\mu
-4\lambda \mu .  \label{Payoffs_factor}
\end{eqnarray}

For the case when $\mu =\nu $ i.e. the players \textrm{B} and \textrm{C}
make a coalition and play the same strategy, whereas the player \textrm{A}
is left out, we have

\begin{eqnarray}
\Pi _{\mathrm{A}}(\lambda ,\mu ,\mu ) &=&2(\lambda -\mu )(2\mu -1),  \notag
\\
\Pi _{\mathrm{B}}(\lambda ,\mu ,\mu ) &=&\Pi _{\mathrm{C}}(\lambda ,\mu ,\mu
)=(\lambda -\mu )(1-2\mu ).
\end{eqnarray}%
For the case when $\lambda =\nu $ i.e. the players \textrm{A} and \textrm{C}
make a coalition and play the same strategy, whereas the player \textrm{B}
is left out, we have

\begin{eqnarray}
\Pi _{\mathrm{A}}(\lambda ,\mu ,\lambda ) &=&(\lambda -\mu )(2\lambda
-1)=\Pi _{\mathrm{C}}(\lambda ,\mu ,\lambda ),  \notag \\
\Pi _{\mathrm{B}}(\lambda ,\mu ,\lambda ) &=&2(\mu -\lambda )(2\lambda -1).
\end{eqnarray}%
For the case when $\lambda =\mu $ i.e. the players \textrm{A} and \textrm{B}
make a coalition and play the same strategy whereas the player \textrm{C} is
left out, we have

\begin{eqnarray}
\Pi _{\mathrm{A}}(\lambda ,\lambda ,\nu ) &=&(\lambda -\nu )(2\lambda
-1)=\Pi _{\mathrm{B}}(\lambda ,\lambda ,\nu ),  \notag \\
\Pi _{\mathrm{C}}(\lambda ,\lambda ,\nu ) &=&2(\nu -\lambda )(2\lambda -1).
\end{eqnarray}

In view of these payoff relations, let $\mu =\nu =c,$ i.e. the players 
\textrm{B} and \textrm{C} make a coalition and play the strategy $c$,
whereas the player \textrm{A} is left out and plays $\lambda =l\neq c$,
resulting in

\begin{eqnarray}
\Pi _{\mathrm{B}}(l,c,c) &=&(l-c)(1-2c)=\Pi _{\mathrm{C}}(l,c,c), \\
\Pi _{\mathrm{A}}(l,c,c) &=&-2(l-c)(1-2c)=-2\Pi _{\mathrm{B}}(l,c,c),
\end{eqnarray}%
i.e. when $l=c$ we get $\Pi _{\mathrm{A}}(l,c,c)=0$, as expected. Also when $%
l=1-c$ we obtain

\begin{equation}
\Pi _{\mathrm{A}}(l,c,c)=-2(2c-1)^{2}=\Pi _{\mathrm{B}}(c,l,c)=\Pi _{\mathrm{%
C}}(c,c,l),
\end{equation}%
That is, the coalition can be considered as a single player. Noticing $\Pi _{%
\mathrm{B}}(l,c,c)=l-2cl-c+2c^{2}$ we now consider

\begin{eqnarray}
\Pi _{\mathrm{A}}(l^{\ast },c^{\ast },c^{\ast })-\Pi _{\mathrm{A}}(l,c^{\ast
},c^{\ast }) &=&\frac{\partial \Pi _{\mathrm{A}}}{\partial l}\mid _{(l^{\ast
},c^{\ast })}(l^{\ast }-l)=-2(1-2c^{\ast })(l^{\ast }-l)\geq 0,  \notag \\
\Pi _{\mathrm{B}}(l^{\ast },c^{\ast },c^{\ast })-\Pi _{\mathrm{B}}(l^{\ast
},c,c) &=&\frac{\partial \Pi _{\mathrm{B}}}{\partial c}\mid _{(l^{\ast
},c^{\ast })}(c^{\ast }-c)=(-2l^{\ast }-1+4c^{\ast })(c^{\ast }-c)\geq 0, 
\notag \\
&&
\end{eqnarray}%
that gives $c^{\ast }=l^{\ast }=1/2.$

\subsection{Quantum solution}

Using (\ref{Conditions}), the players' mixed-strategy payoff relations (\ref%
{mixed_strategy_payoffs}) can be re-expressed as

\begin{eqnarray}
\Pi _{\mathrm{A}}(\lambda ,\mu ,\nu )
&=&2[P(ab)-2P(bc)+P(ac)-P(a)]+P(b)+P(c),  \notag \\
\Pi _{\mathrm{B}}(\lambda ,\mu ,\nu )
&=&2[P(ab)+P(bc)-2P(ac)-P(b)]+P(a)+P(c),  \notag \\
\Pi _{\mathrm{C}}(\lambda ,\mu ,\nu )
&=&2[-2P(ab)+P(bc)+P(ac)-P(c)]+P(a)+P(b).  \label{Payoffs_cooperative}
\end{eqnarray}%
Now, with the pure state (\ref{Pure_state}) these payoffs (\ref%
{Payoffs_cooperative}) take the form

\begin{eqnarray}
\Pi _{\mathrm{A}}(\lambda ,\mu ,\nu ) &=&2[\left\vert c_{2}\right\vert
^{2}+\left\vert c_{3}\right\vert ^{2}-2\left\vert c_{4}\right\vert
^{2}-2\left\vert c_{5}\right\vert ^{2}+\left\vert c_{6}\right\vert
^{2}+\left\vert c_{7}\right\vert ^{2}]-2\lambda +\mu +\nu ,  \notag \\
\Pi _{\mathrm{B}}(\lambda ,\mu ,\nu ) &=&2[\left\vert c_{2}\right\vert
^{2}-2\left\vert c_{3}\right\vert ^{2}+\left\vert c_{4}\right\vert
^{2}+\left\vert c_{5}\right\vert ^{2}-2\left\vert c_{6}\right\vert
^{2}+\left\vert c_{7}\right\vert ^{2}]+\lambda -2\mu +\nu ,  \notag \\
\Pi _{\mathrm{C}}(\lambda ,\mu ,\nu ) &=&2[-2\left\vert c_{2}\right\vert
^{2}+\left\vert c_{3}\right\vert ^{2}+\left\vert c_{4}\right\vert
^{2}+\left\vert c_{5}\right\vert ^{2}+\left\vert c_{6}\right\vert
^{2}-2\left\vert c_{7}\right\vert ^{2}]+\lambda +\mu -2\nu .
\end{eqnarray}%
Consider now the following constraints on the three-qubit pure quantum state

\begin{equation}
\left\vert c_{4}\right\vert ^{2}=\left\vert c_{6}\right\vert ^{2}=\left\vert
c_{7}\right\vert ^{2},\text{ }\left\vert c_{2}\right\vert ^{2}=\left\vert
c_{3}\right\vert ^{2}=\left\vert c_{5}\right\vert ^{2},  \label{Coop-cond}
\end{equation}%
which reduce the above payoffs to

\begin{eqnarray}
\Pi _{\mathrm{A}}(\lambda ,\mu ,\nu ) &=&-2\lambda +\mu +\nu ,  \notag \\
\Pi _{\mathrm{B}}(\lambda ,\mu ,\nu ) &=&\lambda -2\mu +\nu ,  \notag \\
\Pi _{\mathrm{C}}(\lambda ,\mu ,\nu ) &=&\lambda +\mu -2\nu .
\end{eqnarray}%
Now the definitions (\ref{Marginals_Coeffs}) makes the players' strategic
variables $\lambda ,$ $\mu ,$ and $\nu $ to become identical and also that
each player's payoff then become identically zero i.e. $\Pi _{\mathrm{A}%
}(\varkappa )=\Pi _{\mathrm{B}}(\varkappa )=\Pi _{\mathrm{C}}(\varkappa )=0.$
This results in the players left with no motivation to make a coalition.

\section{Discussion}

Fine's theorem establishes a connection between the existence of a joint
probability distribution for specified marginals and the satisfaction of
system of Bell's inequalities, linked through the \textit{if and only if}
conditions. Using Fine's theorem, our analysis rephrases the mixed-strategy
payoff relations of a game in the context of the marginals of a joint
probability distribution. If such a joint distribution exists for the
marginals, its depiction in terms of these marginals can be further
simplified to an expression involving the joint distribution itself.
Crucially, if this joint distribution is factorizable, the game can be
distilled down to its classical mixed-strategy counterpart, thus making the
mixed-strategy classical game to remain embedded within the presented
framework. That is, the mere existence of a joint distribution corresponding
to a set of marginals does not necessarily yield the classical
mixed-strategy game. Factorizability of this distribution, in the sense
described, remains a key requisite. That is, a factorizable probability
distribution---as outlined---leads to the classical mixed-strategy game,
ensuring that the set of Bell's inequalities is met. However, it's worth
noting that a probability distribution satisfying Bell's inequalities
doesn't necessarily guarantee its factorizability in the described manner.

Furthermore, through the lens of Fine's theorem, we demonstrate that player
payoffs and strategies---when re-expressed in terms of a set of
marginals---set the stage for analyzing what quantum probabilities can offer
in potentially resolving the dilemmas that the games can represent. Via
Fine's theorem, along with the considered or given quantum state(s) and the
mixed-strategy version of the classical game, the presented framework allows
a direct consideration of what quantum probabilities can offer to the game.
In order to obtain required marginal from the quantum states, the framework
uses POVMs.

When applied to three-player PD, a three-qubit pure state (\ref{PD-state})
is determined for which the presented framework gives an unattractive
outcome, with a continuum of NE that emerge. When applied to a three-player
symmetric cooperative game, sets of three-qubit pure states, as given by (%
\ref{Coop-cond}), are determined for which the players are left with no
motivation to make a coalition. The framework paves the way for the analysis
of a range of games from the perspective of what quantum states have to
offer and that can resolve their inherent dilemmas.

The presented approach to quantization is \textit{complete} in the sense
defined in Ref. \cite{Bleiler2008}, as the mixed-strategy version of the
classical game remains embedded within its quantized version. Ref. \cite%
{Kolokotsov2020} describes how POVMs generalize quantization schemes based
on EWL \cite{EWL1,EWL2} and MW \cite{MarinattoWeber} schemes. The present
manuscript, however, uses POVMs more directly in view of applying Fine's
theorem in order to re-express the player payoffs and strategies in term of
a set of marginals.

\section{Conclusions}

In essence, the introduced methodology paves the way for obtaining a game's
quantum version directly from its classical mixed-strategy iteration,
keeping in perspective the existence and/or non-existence of a joint
probability distribution corresponding to a set of marginals, corresponding
system of Bell's inequalities, and Fine's theorem that links them. The
proposed framework is not confined to Nashian game theory. Instead, it
offers the versatility to incorporate non-Nashian game-theoretic solution
concepts. This encompasses a range of concepts such as perfect prediction
equilibrium, translucent equilibrium, perfectly transparent equilibrium,
superrationality, and minimax-rationalizability, among others. Furthermore,
an intuitive progression of this research would be to apply Fine's theorem
to four bivalent observables \cite{Fine,Fine1,Fine2}, paving the way for the
quantization of four-player games.

\end{document}